\documentclass[10pt,journal,cspaper,compsoc]{IEEEtran}  

\usepackage[pdftex]{graphicx}
 \usepackage[bookmarks=false]{hyperref}
\usepackage{url}
\usepackage{multirow}
\usepackage[table]{xcolor}
\usepackage{algorithm}
\usepackage{algpseudocode}
\usepackage[latin1]{inputenc}

\let\labelindent\relax

\usepackage{enumitem}
\usepackage{subfigure}
\usepackage{pdfcomment}

\begin{document}
%
\title{Cloud Benchmarking For Maximising Performance of Scientific Applications}

\author{Blesson Varghese, 
        Ozgur Akgun, 
        Ian Miguel,  
        Long Thai and
        Adam Barker  
        
\IEEEcompsocitemizethanks{ \IEEEcompsocthanksitem 
E-mail: varghese@qub.ac.uk, \{ozgur.akgun, ijm, ltt2, adam.barker\}@st-andrews.ac.uk 
}
\thanks{}}

\markboth{IEEE Transactions on Cloud Computing,~Accepted 31 Jul~2016}%
{1 \MakeLowercase{\textit{Varghese et al.}}: Cloud Benchmarking For Maximising Performance of Scientific Applications}


\IEEEcompsoctitleabstractindextext{%
\begin{abstract}
\textit{How can applications be deployed on the cloud to achieve maximum performance?} This question is challenging to address with the availability of a wide variety of cloud Virtual Machines (VMs) with different performance capabilities. The research reported in this paper addresses the above question by proposing a six step benchmarking methodology in which a user provides a set of weights that indicate how important memory, local communication, computation and storage related operations are to an application. The user can either provide a set of four abstract weights or eight fine grain weights based on the knowledge of the application. The weights along with benchmarking data collected from the cloud are used to generate a set of two rankings - one based only on the performance of the VMs and the other takes both performance and costs into account. The rankings are validated on three case study applications using two validation techniques. The case studies on a set of experimental VMs highlight that maximum performance can be achieved by the three top ranked VMs and maximum performance in a cost-effective manner is achieved by at least one of the top three ranked VMs produced by the methodology. 
\end{abstract}

}

\maketitle

\IEEEdisplaynotcompsoctitleabstractindextext


\IEEEpeerreviewmaketitle

\section{Introduction}
\label{introduction}
\IEEEPARstart{T}{he} cloud computing marketplace offers a wide variety of on-demand resources with a wide range of performance capabilities. This makes it challenging for a user to make an informed choice as to which Virtual Machines (VMs) need to be selected in order to deploy an application to achieve maximum performance. Often it is the case that users deploy applications on an ad hoc basis, without understanding which VMs can provide maximum performance. This can result in the application under performing on the cloud and consequently increasing running costs. This paper aims to address the above problem. 

The above problem is addressed by benchmarking to measure the performance of computing resources \cite{benchmark-1, benchmark-2}, which is employed on the cloud \cite{cloudbenchmark-0, cloudbenchmark-1, cloudbenchmark-2}. Typically, cloud benchmarking is performed independently of the application that needs to be deployed, and does not consider bespoke requirements an application might have. Our research advances the start-of-the-art by proposing an application aware benchmarking methodology that accounts for the requirements of the applications.  

We hypothesise that by taking into account the requirements of an application, along with benchmarking data collected from the cloud, VMs can be ranked in order of performance and cost effectiveness so that a user can deploy an application on a cloud VM, which will maximise performance. In this paper, the focus is on scientific High-Performance Computing (HPC) applications and maximum performance is defined as the minimum execution time of an application. Our motivation for choosing scientific applications is that the cloud is becoming an alternate computing platform to do HPC without owning supercomputers \cite{cloudhpc-1, cloudhpc-2}. However, on the cloud, applications will need to use a pay-as-you-go model for running on cloud VMs that share physical nodes with other applications. This is in contrast to grants and quotas which are available for utilising tightly coupled nodes of supercomputers which are less likely to be shared with other applications. Deploying long running applications in an ad hoc manner on the cloud will result in under performance and increased running costs. In this context, a method to determine cloud VMs that can maximise the performance of the application before it is deployed is required given the wide variety of choices offered by providers to do HPC on the cloud.

We present a six step benchmarking methodology in order to determine the VMs that can maximise the performance of scientific applications on the cloud. A user provides as input a set of weights that describe the memory and process, local communication, computation, and storage requirements of the scientific application to be deployed on the cloud. These requirements are mapped onto a set of four aggregate groups or eight fine grain groups that capture the memory and process, local communication, computation, and storage attributes of the VMs. The groups are obtained by benchmarking the cloud VMs. Based on the user's knowledge of the application (developers and domain experts are well acquainted with their applications), either a set of four abstract weights or eight fine grain weights are provided as input. The value of each weight ranges from 0 to 5, where 0 signifies that the memory and process, local communication, computation, or storage groups represented by the weight has no relevance to the application, and 5 indicates that the group is important to the application for achieving maximum performance. These weights along with the benchmarked data are used to generate rankings that take performance and cost into account. 


For the purposes of verifying our hypothesis, the methodology is validated on three scientific HPC applications; the first is used in financial risk, the second employed in molecular dynamics, and the third as a mathematical solver. The memory and process, local communication, computation and storage requirements of these applications are known beforehand. The applications are embarrassingly parallel and in this research we parallelise them on the multiple cores of a single VM. Two techniques, namely comparative validation and enumeration-based validation are used to validate the rankings produced by the methodology. The validation study demonstrates that the methodology can select cloud VMs that maximise the performance of the application. If such a methodology is not adopted, the application will result in higher running costs. The contributions of this paper are the development of a benchmarking methodology for selecting VMs that maximises the performance of scientific applications on the cloud, and the validation of the methodology against real world applications. 

The remainder of this paper is organised as follows. Section \ref{cloudbenchmarking} presents the cloud benchmarking methodology proposed in this paper for maximising an application's performance on the cloud. Section \ref{cloudvirtualmachines} considers the Amazon Elastic Compute Cloud (EC2) VMs employed in this paper. Section \ref{benchmarks} presents the  benchmarks used in the methodology. 
Section \ref{cloudrankings} presents the aggregate and fine-grain cloud rankings generated by the methodology for different weights. 
Section \ref{sensitivityanalysis} considers three case study applications to validate the cloud benchmarking methodology. Section \ref{relatedwork} presents a discussion on the work related to the research presented in this paper. Section \ref{conclusions} concludes this paper. 

\section{Cloud Benchmarking Methodology}
\label{cloudbenchmarking}

A six step cloud benchmarking methodology that determines which cloud Virtual Machines (VMs) can maximise application performance on the cloud is proposed. The six steps are: (1) capture attributes of cloud VMs, (2) group attributes of cloud VMs, (3) benchmark cloud VMs, (4) normalise attribute groups, (5) provide weights to groups, and (6) rank cloud VMs. Only the last two steps are necessary to be performed for each application deployment on the cloud. The first four steps are performed infrequently (only if the underlying infrastructure has changed in a private cloud or periodically for a public cloud). 


Individual attributes of cloud VMs are firstly evaluated and then grouped together. 
The user provides a set of weights (or the order of importance) for the groups based on the knowledge of the requirements of the application to be deployed on the cloud. The weights along with the attributes of the group for each VM are used to generate a score resulting in a VM rank. Two sets of ranks are generated; the first ranking is solely based on the performance of the VMs and the second ranking considers both performance and cost. We hypothesise from the first set of ranks that the VMs with the highest ranks can maximise application performance on the cloud, and the highest ranks from the second set can maximise application performance in a cost-effective manner. 

Given $i = 1, 2, \cdots , m$ different VMs, the cloud benchmarking methodology we propose is as follows:

\subsubsection*{Step 1: Capture Attributes}
The attributes of a VM that describes it are firstly selected based on experience with VMs and physical machines. For example, (a) attributes such as the number of integer, float and double addition, multiplication and division operations that can be performed in one second on a VM can describe its computational capacity, or (b) attributes such as the number of sequential and random read, write and delete operations that can be performed in one second on the storage of a VM can describe its file I/O. Assume that there are $j = 1, 2, \cdots , n$ attributes of a VM. Then, $r_{i,j}$ is the value associated with the $j^{th}$ attribute on the $i^{th}$ VM. 

\subsubsection*{Step 2: Group Attributes}
The attributes of the VM are then grouped into categories based on whether they are related to memory and process, local communication, computation or storage. For example, (a) attributes such as the bandwidth of memory read and write operations and of communication using pipes, AF Unix socket and TCP are grouped together as the local communication group, or (b) attributes related to the latencies of the main and random access memory and the L1 and L2 cache can be grouped as the memory group. Each attribute group is denoted as 
$G_{i, k} = \{r_{i, 1}, r_{i, 2}, \cdots \}$, where $i = 1, 2, \cdots m$, $k = 1, 2, \cdots , p$, and $p$ is the number of attribute groups.

\subsubsection*{Step 3: Benchmark Virtual Machines}
Based on the attribute groups a set of benchmarks are evaluated on all potential cloud VMs. The benchmarks evaluate the attributes of the cloud VMs as closely as possible to the underlying hardware employed \cite{benchmark-2, cloudbenchmark-80, cloudbenchmark-81}. The benchmarks are grouped as memory and process, local communication, computation and storage evaluation groups, based on \textit{Step 2}. Standard benchmark tools are run on the cloud VM or on an observer system collecting the results. The focus is on evaluating attributes that are closer to the hardware, such as frequencies (number of operations in a second) latencies (micro seconds or nano seconds) and bandwidths (Megabytes per second). The value of each attribute, $r_{i, j}$ is obtained in this step. 

\subsubsection*{Step 4: Normalise Groups}
The value of the attributes are normalised to rank the performance of VMs for an attribute group. The group based rank can provide a view of the relative performance of the VM under each group. For example, a VM that is well ranked in the memory group may poorly perform in the storage group. The normalised value \cite{normalise} of each attribute $\bar{r}_{i, j} = \frac{r_{i, j} - \mu_{j}}{\sigma_{j}}$, where $\mu_j$ is the mean value of attribute $r_{i, j}$ over $m$ VMs (the mean value is the sum of all values of an attribute for $m$ VMs and dividing that sum by $m$) and $\sigma_j$ is the standard deviation of the attribute $r_{i, j}$ over $m$ VMs (the standard deviation is the sum of the squares of the difference between the value of each attribute and mean for $m$ VMs and dividing that sum by $m$). The resultant normalised attribute group, is denoted as $\bar{G}_{i, k} = \{\bar{r}_{i, 1}, \bar{r}_{i, 2}, \cdots \}$, where $i = 1, 2, \cdots m$, $k = 1, 2, \cdots , p$, and $p$ is the number of attribute groups.

\subsubsection*{Step 5: Weight Groups}
For a given application, some attribute groups may be more important than the others. This is known to domain experts and application developers who are familiar with the application. For example, (a) a financial risk simulation for computing Value-at-Risk and ingests 2 GB of data may not have large storage related requirements but will need efficient computation and large I/O bandwidth between the processor and memory, hence, having a greater weight for the local communication group or (b) for a simulation requiring 500 Gigabyte space, the storage group is relevant. To capture this, after normalising, a weight for each attribute group is provided by the user, which is defined as $W_{k}$. In this paper, $W_{k}$ can take values from 0 to 5, where 0 indicates that the group has no relevance for that application and 5 indicates that the group has the highest importance for the application.  

\subsubsection*{Step 6: Rank Virtual Machines}
The score of each VM is calculated as $S_{i} = \bar{G}_{i, k}.W_{k} = \sum\limits_{i'=1}^{i_{max}}\bar{r}_{i, i'}.W{k}$, where $max$ is the number of attributes in the attribute group. 
For attributes in the group where a higher value should denote a higher score, for example bandwidth or frequency, the weight used is $W{k}$. On the other hand when a lower value should denote a higher score, for example latency, the weight used is $-W{k}$. 

The scores are ordered in a descending order for generating $Rp_{i}$ which is the ranking of the VMs based solely on performance. The performance ranks are used in one of the validation techniques presented in Section \ref{sec:validationtechniques}.

An additional ranking of the VMs, $Rc_{i}$, based on both performance and costs is generated. To obtain this rank, $C_{i} / S_{i}$ are ordered in ascending order, where $C_{i}$ is the cost in \$/hour of the VM. The motivation for including costs in the methodology is determining the highest value-for-money VMs instead of merely finding the best performing VMs.

\section{Cloud Virtual Machines}
\label{cloudvirtualmachines}
The Amazon Elastic Compute Cloud (EC2)\footnote{\url{http://aws.amazon.com/ec2/}} platform is chosen since it is publicly available and offers a variety of VMs with different performance capabilities. Different categories of previous generation EC2 VMs, referred to as instances, are offered as general purpose, memory optimised, cluster compute optimised, storage optimised, and GPU instances\footnote{\url{https://aws.amazon.com/ec2/previous-generation/}}. Table \ref{table1} shows the specification of the underlying hardware of the instances which are used for benchmarking (instances with more than 15 GiB are chosen to facilitate smooth running of the case study applications). The region and availability zones of the instances are US East N. Virginia (us-east-1) and US West Oregon (us-west-2).

\begin{table}[!ht]
\begin{center}
	\caption{Amazon Elastic Compute Cloud (EC2) instances employed for benchmarking}
	\begin{tabular}{p{1.5cm} p{0.6cm} p{0.5cm} p{2.35cm} p{0.5cm} p{0.7cm}}
		\hline	
		\textbf{Instance Type}	&	\textbf{Virtual CPUs (vCPU)}	&	\textbf{Mem. (GiB)}	&	\textbf{Processor}	&			\textbf{Clock Speed (GHz)}	& 	\textbf{Cost/Hr (\$)~\cite{amazonec2instancepricing}}\\
		\hline		
		\texttt{m1.xlarge}	&	4	&	15.0	&	Intel Xeon E5-2650	&	2.00	&	0.480\\
		\hline
		\texttt{m2.xlarge}	&	2	&	17.1	&	Intel Xeon E5-2665	&	2.40	& 0.410\\
		\texttt{m2.2xlarge}	&	4	&	34.2	&	Intel Xeon E5-2665	&	2.40	& 0.820\\
		\texttt{m2.4xlarge}	&	8	&	68.4	&	Intel Xeon E5-2665	& 	2.40	& 1.640\\
		\hline
		\texttt{m3.xlarge}	&	4	&	15.0	&	Intel Xeon E5-2670	& 	2.60	& 0.500\\
		\texttt{m3.2xlarge}	&	8	&	30.0	&	Intel Xeon E5-2670	& 	2.60	& 1.000\\
		\hline
		\texttt{hi1.4xlarge}&	16	&	60.5	&	Intel Xeon E5620	&	2.40	&	3.500\\
		\texttt{hs1.4xlarge}&	16	&	117.0	&	Intel Xeon E5-2650	&	2.00	&	4.600\\
		\hline
		\texttt{cc1.4xlarge}&	16	&	23.0	&	Intel Xeon X5570	&	2.93	&	1.300\\
		\texttt{cc2.8xlarge}&	32	&	60.5	&	Intel Xeon X5570	&	2.93	&	2.400\\
		\hline		
		\texttt{cr1.4xlarge}&	32	&	244.0	&	Intel Xeon E5-2670	& 	2.60	&	3.500\\		
		\hline
		\hline		
	\end{tabular}
	\label{table1}
	\end{center}
\end{table}

The general purpose instances are \texttt{m1} and \texttt{m3} instances, the memory optimised instances are \texttt{m2} and \texttt{cr1} instances, the compute optimised instances are \texttt{cc1} and \texttt{cc2} and the storage optimised instances are \texttt{hi1} and \texttt{hs1}. All instances are abstracted over their respective processors. Each virtual CPU (vCPU) of the \texttt{m3}, \texttt{cr1} and \texttt{cc2} instances is a hyperthread on a core of the underlying processor.

\section{Cloud Benchmarks}
\label{benchmarks}
The experimental setup for obtaining the attributes of VMs by benchmarking and then grouping them are presented in this section. The attributes $r_{i, j}$ of Step 3 in Section \ref{cloudbenchmarking} are obtained and then grouped to obtain $G_{i, k}$.   

\subsection{Setup}
Three tools, namely \texttt{bonnie++}, \texttt{lmbench} and \texttt{sysbench} are employed for benchmarking. The \texttt{bonnie++}\footnote{\url{http://sourceforge.net/projects/bonnie/}} tool is used for file system benchmarks. The time and latency for reading data from file and writing data to file, sequential and random create, read and write operations, and number of seeks performed in a second can be benchmarked. 

The \texttt{lmbench}\footnote{\url{http://lmbench.sourceforge.net/}} tool provides latency and bandwidth information on top of a wide range of memory and process related information \cite{lmbench-1}. Context switching times and VMs latencies are also obtained from this tool.

The \texttt{sysbench}\footnote{\url{http://sysbench.sourceforge.net/}} tool, commonly referred to as the Multi-threaded System Evaluation Benchmark, is also used for obtaining benchmark metrics related to the CPU and the file I/O performance. It is popular for understanding a system under data intensive loads. 

All experiments to gather the benchmark metrics considered in the following section were performed eight times consecutively during a seven week period in September and October 2013. The instances with over 23 GiB memory could not be easily benchmarked. This is because file sizes used for benchmarking need to be at least twice as large as the memory to avoid caching of the file in memory which produces incorrect benchmark metrics. For this reason the size of the files used for benchmarking were up to 750 GiB nearly times the size of the highest memory for \texttt{bonnie++} and \texttt{sysbench} to obtain more accurate results. A number of instances cannot accommodate large files due to the limited storage available. Additional Amazon Elastic Block Storage (EBS)\footnote{\url{http://aws.amazon.com/ebs/}} volumes had to be attached to the instances to successfully benchmark the instances. 

Table \ref{benchmarkcosttable1} shows the time and costs incurred for running the benchmarks on different VMs. This highlights the overhead not only on the time spent for obtaining the benchmarks but also on the total costs for using the resources, including the instances and storage. 

\begin{table}
	\centering
	\caption{Time and cost for executing benchmarks}
	\begin{tabular}{| c | p{2.5cm} | p{2cm} |}
	\hline
	\textbf{VM}	& \textbf{Time (hrs) $\times$ Cost (\$/hour)}	&	\textbf{Cost (\$) for 8 executions} \\
	\hline
	\hline
	m1.xlarge		&	2.5 $\times$ 0.48	&	9.6\\
	m2.xlarge		&	2.5 $\times$ 0.41	&	8.2\\
	m2.2xlarge		&	5.0 $\times$ 0.82	&	32.8\\
	m2.4xlarge		&	10.0 $\times$ 1.64	&	131.2\\
	m3.xlarge		&	2.0 $\times$ 0.5	&	8.0\\
	m3.2xlarge		&	2.5 $\times$ 1.0	&	20.0\\
	hi1.4xlarge		&	7.5 $\times$ 3.5	&	210.0\\
	hs1.8xlarge		&	14.0 $\times$ 4.6	&	515.2\\
	cc1.4xlarge		& 	5.0 $\times$ 1.3	&	52.0\\
	cc2.8xlarge		&	6.0 $\times$ 2.4	&	115.2\\
	cr1.4xlarge		&	10.0 $\times$ 3.5	&	280.0\\
	\hline
	\end{tabular}
	\label{benchmarkcosttable1}
\end{table}

\subsection{Attributes and Groups}
\begin{figure*}[htp]
\begin{center}
	\subfigure[Memory Latencies: L1 and L2 cache] {\includegraphics[width=0.485\textwidth]{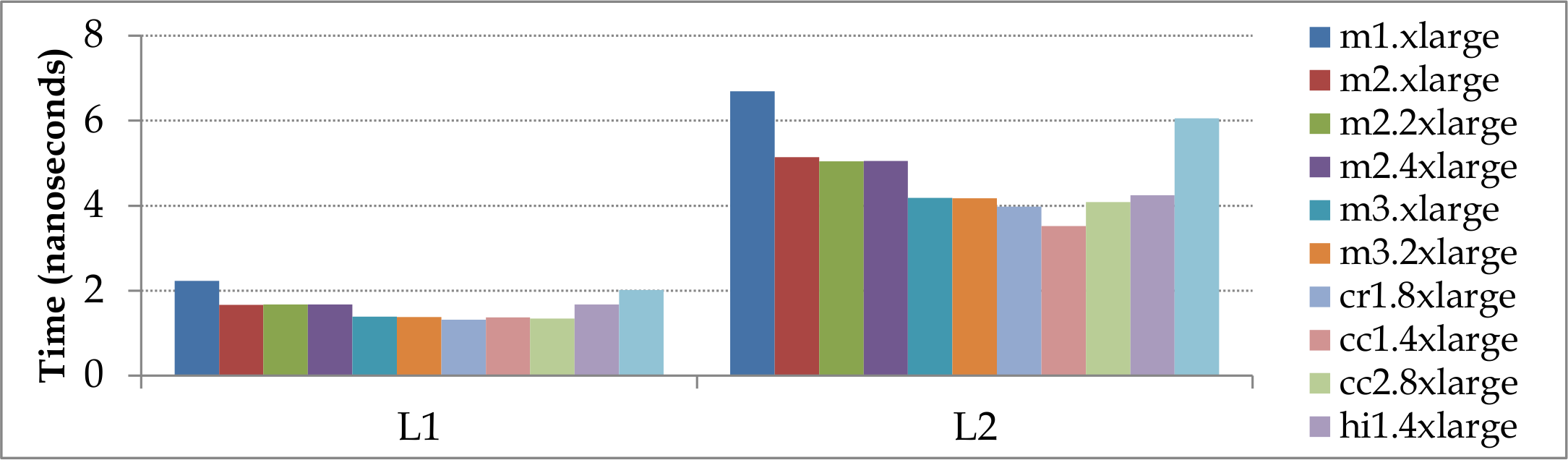} \label{figure3-1-1}}
	\subfigure[Memory Latencies: Main and Random Memory]{\includegraphics[width=0.485\textwidth]{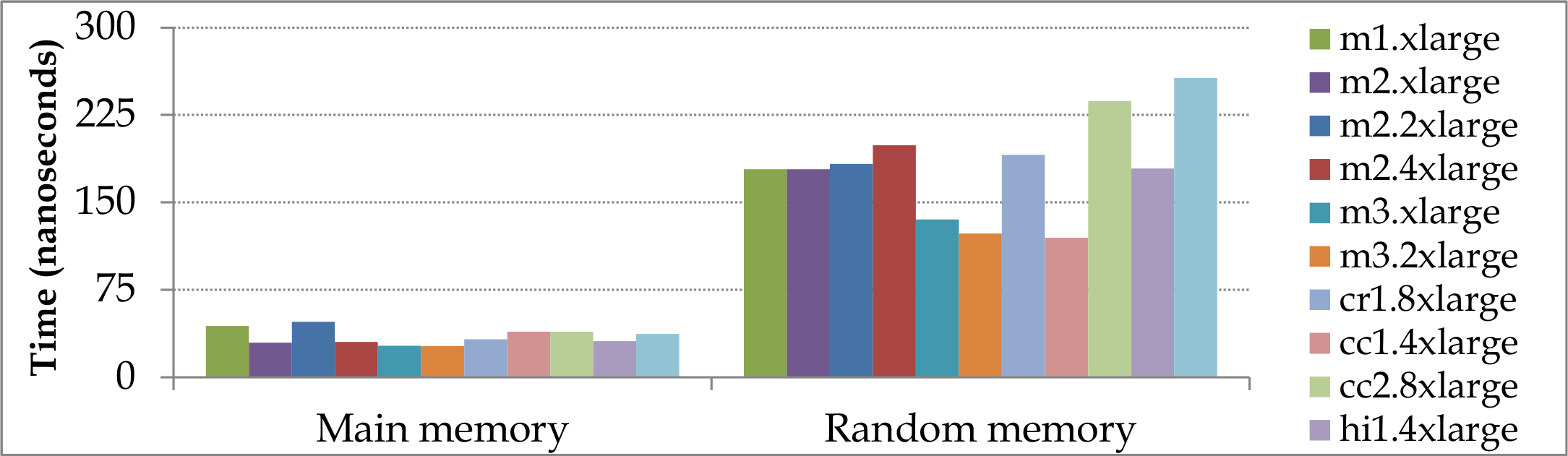} \label{figure3-1-2}}

	\subfigure[Process Latencies: Context Switching]{\includegraphics[width=0.99\textwidth]{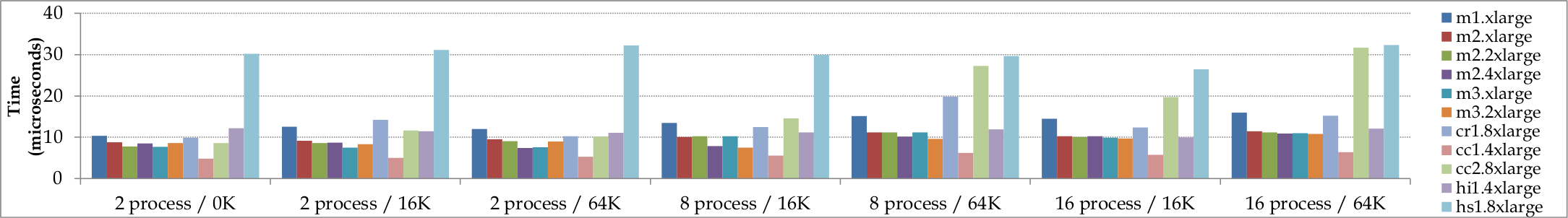} \label{figure3-1-3}}
	\subfigure[Local communication bandwidth]{\includegraphics[width=0.99\textwidth]{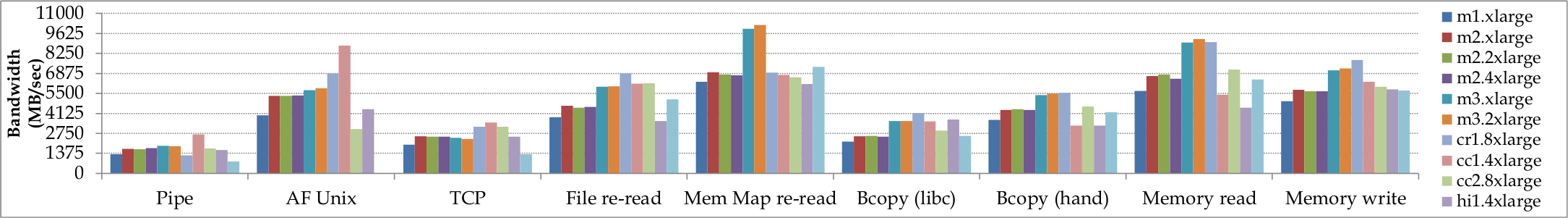} \label{figure3-2-1}}

	\subfigure[Arithmetic Operation Time: Addition and multiplication]{\includegraphics[width=0.535\textwidth]{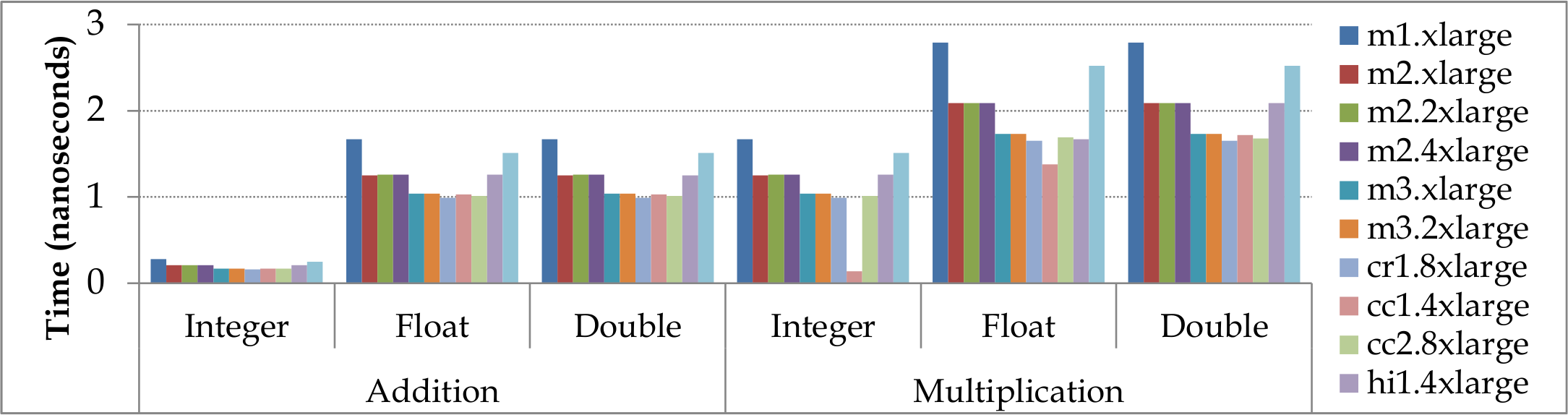} \label{figure3-3-1}}
	\subfigure[Arithmetic Operation Time: Division and Modulus]{\includegraphics[width=0.432\textwidth]{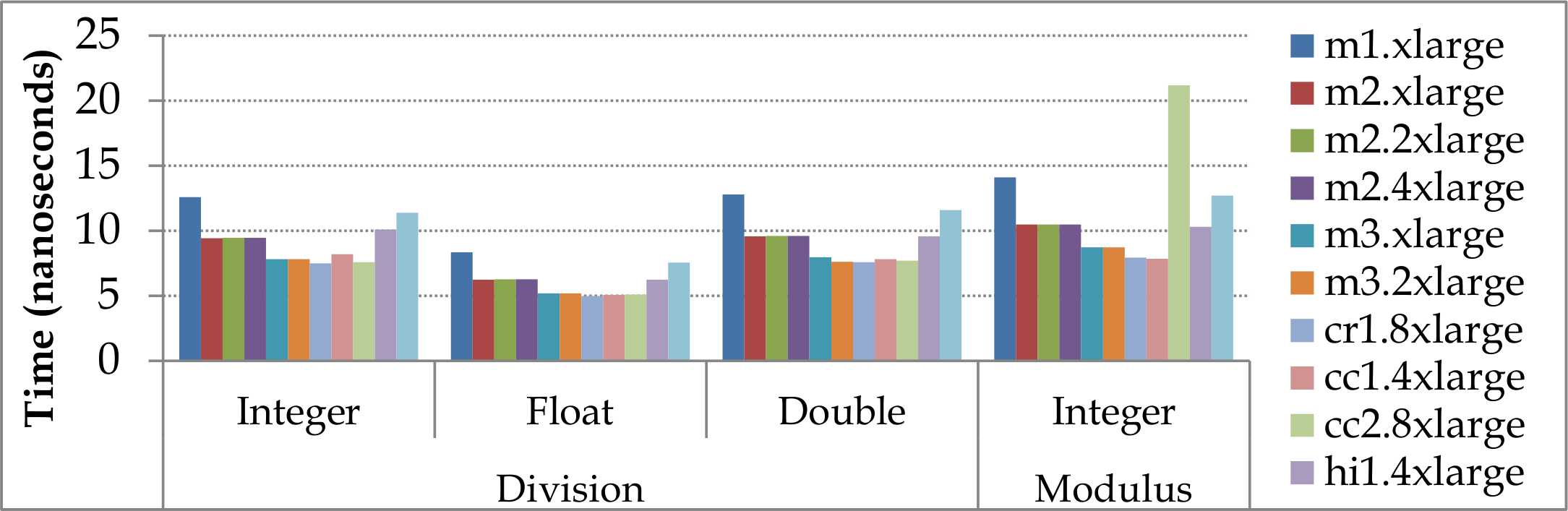} \label{figure3-3-2}}

	\subfigure[File I/O Operations in one Second: Sequential and random create and delete]{\includegraphics[width=0.59\textwidth]{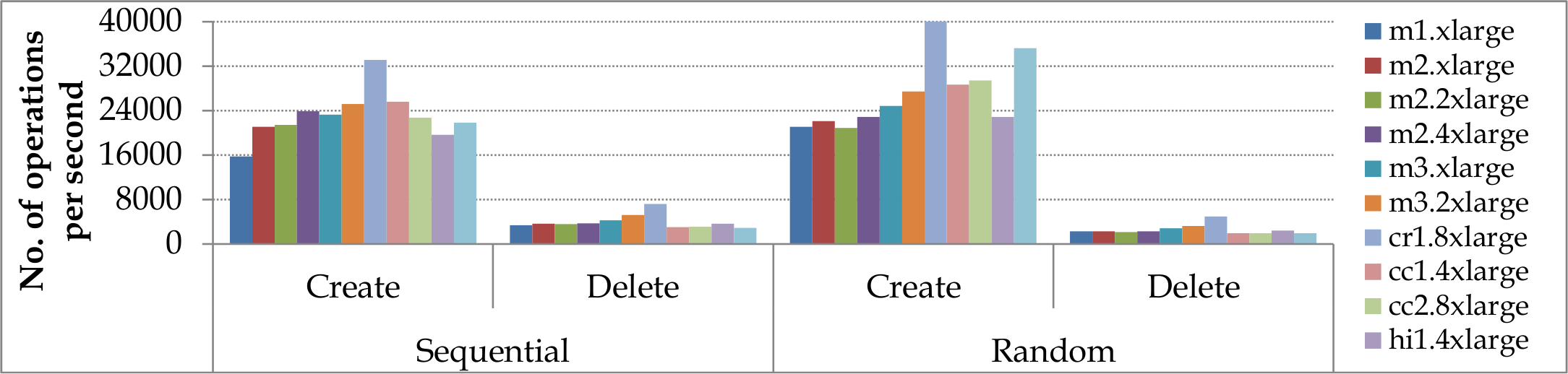} \label{figure3-4-1}}
	\subfigure[File I/O Operations in one Second: Sequential and random read]{\includegraphics[width=0.38\textwidth]{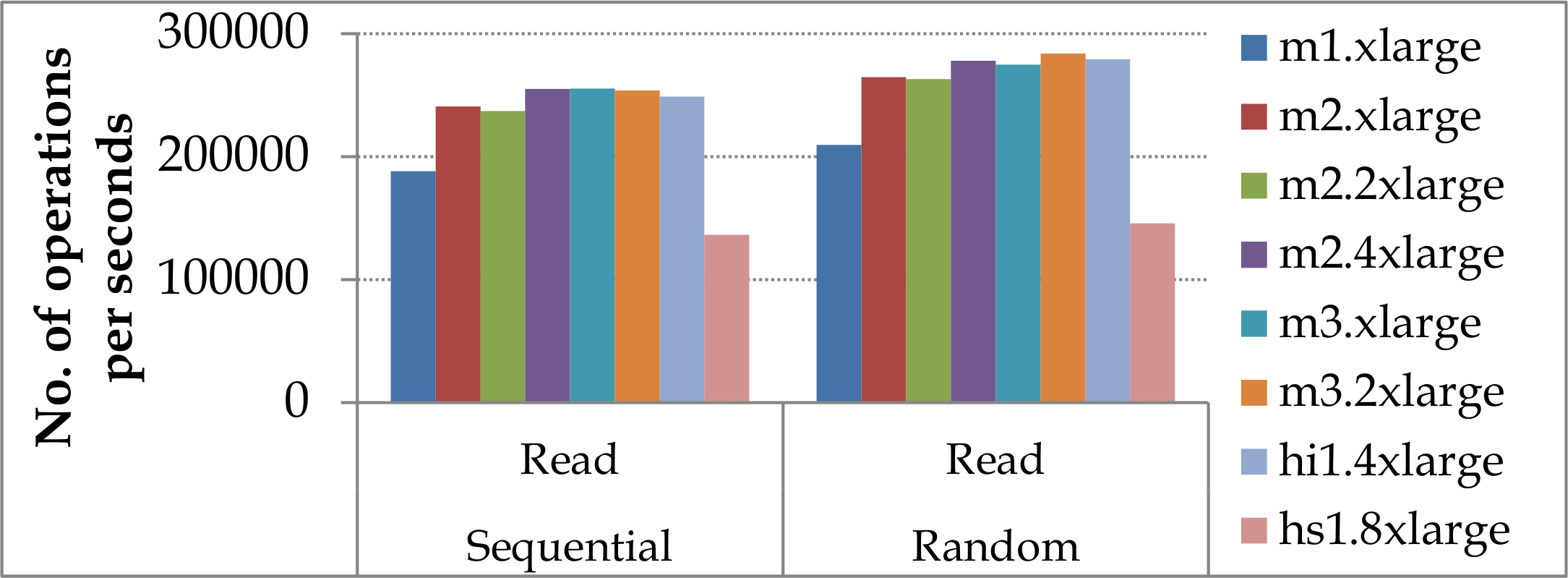} \label{figure3-4-2}}
\caption{Sample benchmarks obtained from 11 Amazon Cloud VMs. Low value of latencies and operation time and high value of bandwidth and operations in one second indicates good performance of VMs.}
\label{fig:benchmarks}
\end{center}
\end{figure*}

Two groupings are used to combine the attributes gathered from the cloud. The first provides an aggregated view of the attributes, whereas the second provides a more detailed view of the benchmarks as shown in Table \ref{grouptable1}. The aggregate set $G_{agg}$, comprises four groups, denoted as $G_{agg} = \{G_1, G_2, G_3, G_4\}$. The fine grain set $G_{sub}$, in which each aggregate group is decomposed into two sub-groups, is denoted as $G_{sub} = \{G_{1,1}, G_{1,2}, G_{2,1}, G_{2,2}, G_{3,1}, G_{3,2}, G_{4,1}, G_{4,2}\}$. Figure \ref{fig:benchmarks} shows the values obtained for a set of sample attributes. A low value of latencies and operation time and a high value of bandwidth and operations in one second indicates good performance of a VM.


\begin{table}
	\centering
	\caption{Aggregate and Fine-grain Groups}
	\begin{tabular}{| c | c | p{4.5cm} |}
	\hline
	\textbf{Group}	& \multicolumn{2}{ c |}{\textbf{Description}}\\
	\hline
	\hline
	\multirow{3}{*}{$G_{1}$}	&	\multicolumn{2}{ c |}{Memory and process}\\
	\cline{2-3}
	&	$G_{1,1}$	&	Process latency\\
	&	$G_{1,2}$	&	Memory latency\\
	\hline
	\multirow{3}{*}{$G_{2}$}	&	\multicolumn{2}{ c |}{Local communication}\\
	\cline{2-3}				
	&	$G_{2,1}$	&	Local communication latency\\
	&	$G_{2,2}$	&	Local communication bandwidth\\
	\hline			
	\multirow{3}{*}{$G_{3}$}	&	\multicolumn{2}{ c |}{Computation}\\
	\cline{2-3}				
	&	$G_{3,1}$	&	Integer operations\\
	&	$G_{3,2}$	&	Floating point operations\\
	\hline			
	\multirow{3}{*}{$G_{4}$}	&	\multicolumn{2}{ c |}{Storage}\\
	\cline{2-3}				
	&	$G_{4,1}$	&	File I/O bandwidth\\
	&	$G_{4,2}$	&	File I/O frequency\\
	\hline
	\end{tabular}
	\label{grouptable1}
\end{table}

\subsubsection{Memory and Process Group}
This group, denoted as $G_{1}$ captures the performance and latencies of the memory and process related operations. The L1 and L2 cache latencies of the instances are shown in Figure \ref{figure3-1-1} and the main memory and random memory latencies of the instances are shown in Figure \ref{figure3-1-2}. 

Context switching is computationally expensive and with tens or hundreds of switches per second a substantial cost is added to the processor. The latencies for context switching 2, 8 and 16 processes with 16 and 64 kilobyte sizes are shown in Figure \ref{figure3-1-3} (For example, in the figure, 2 process / 64K indicates that the latency of context switching two processes is noted and the size of each process is 64K\footnote{\url{http://www.bitmover.com/lmbench/lat\_ctx.8.html}})



The aggregate group is divided into $G_{1,1}$ and $G_{1,2}$ sub-groups that capture all the process latencies and memory latencies respectively. 

\subsubsection{Local Communication Group}
The bandwidth of both memory and interprocess communications are captured under the local communication group, denoted as $G_{2}$. Figure \ref{figure3-2-1} shows memory communication metrics, namely the rate (MB/sec) at which data can be read from and written to memory, and interprocess communication metrics, namely the rate of data transfer between pipes and sockets. 

The latencies in local communication are grouped as $G_{2,1}$ and the associated bandwidth under $G_{2,2}$.

\subsubsection{Computation Group}
The attributes captured in this group, denoted as $G_{3}$, are for benchmarking the performance of integer, single precision and double precision float operations such as addition and multiplication (refer Figure \ref{figure3-3-1}) and division and modulus (refer Figure \ref{figure3-3-2}). The computation benchmarks highlight consistently good performance of the \texttt{m3}, \texttt{cg1}, \texttt{cc1} and \texttt{cr1} instances across integer, float and double operations.

The integer operations and floating point operations are grouped into $G_{3,1}$ and $G_{3,2}$ respectively.  

\subsubsection{Storage Group}
File I/O related attributes are grouped as the storage group, denoted as $G_{4}$, which considers the number of sequential create, read and delete and random create, read and delete operations as shown in Figure \ref{figure3-4-1} and Figure \ref{figure3-4-2}. Overall, the best performer given the benchmarks obtained is the \texttt{cr1} instance. The \texttt{m3} instance is not too far behind in the file I/O performance. 


The attributes that describe bandwidth and frequency of the file I/O operations are grouped as $G_{4,1}$ and $G_{4,2}$ respectively.

\section{Cloud Rankings}
\label{cloudrankings}
Given the two group sets $G_{agg}$ and $G_{sub}$ considered in the previous section, a user can provide a weight, $W_{k}$, for each group $G_k$. The set of weights corresponding to $G_{agg}$ is $W_{agg}=\{W_{1}, W_{2}, W_{3}, W_{4}\}$ and the set of weight for $G_{sub}$ is $W_{sub} = \{W_{1,1}, W_{1,2}, W_{2,1}, W_{2,2}, W_{3,1}, W_{3,2}, W_{4,1}, W_{4,2}\}$. The weights assigned by a user represents the importance of a group with respect to an application the user wants to deploy on a cloud instance. Each weight can take values between 0 and 5, where 0 signifies that the group represented by the weight has no relevance to the application and 5 indicates that the group is important to the application for achieving maximum performance. 

Consider for example, an application that is memory intensive requiring limited storage and the user wants to describe the application using a small set of weights. Aggregate weights can be chosen and $W_{1}$ can be set to 5 for representing the relevance of memory group and $W_{4}$ can be set to 0 for describing the irrelevance of the storage group. However, a user may understand the application in detail and opt for the larger weight set to describe the application. Now, if the memory latencies are more important than the process latencies, then $W_{1,2}$ is set to 5 and $W_{1,1}$ may be set to a lower value. $W_{sub}$ is therefore useful in providing a more detailed description of the application to the benchmarking methodology.

Each weight can take six possible values (0-5), and therefore there are $6^4$ and $6^8$ different combination of weights possible for $W_{agg}$ and $W_{sub}$ respectively. However, a set with all zero values is of no real significance, and hence, the total number of different combinations of aggregate weights is $1,295$ and of fine grain weights is $1,679,615$. Using the benchmarking methodology all possible rankings for different weight combinations were generated. Performance (P) and performance-cost (PC) rankings for sequential and parallel execution were considered. When the performance of parallel execution is to be evaluated the methodology takes the number of vCPUs of the instance into account.

\begin{figure}
\centering
	\subfigure[P ranking: sequential execution]{\label{figure50a}\includegraphics[width=0.46\textwidth]{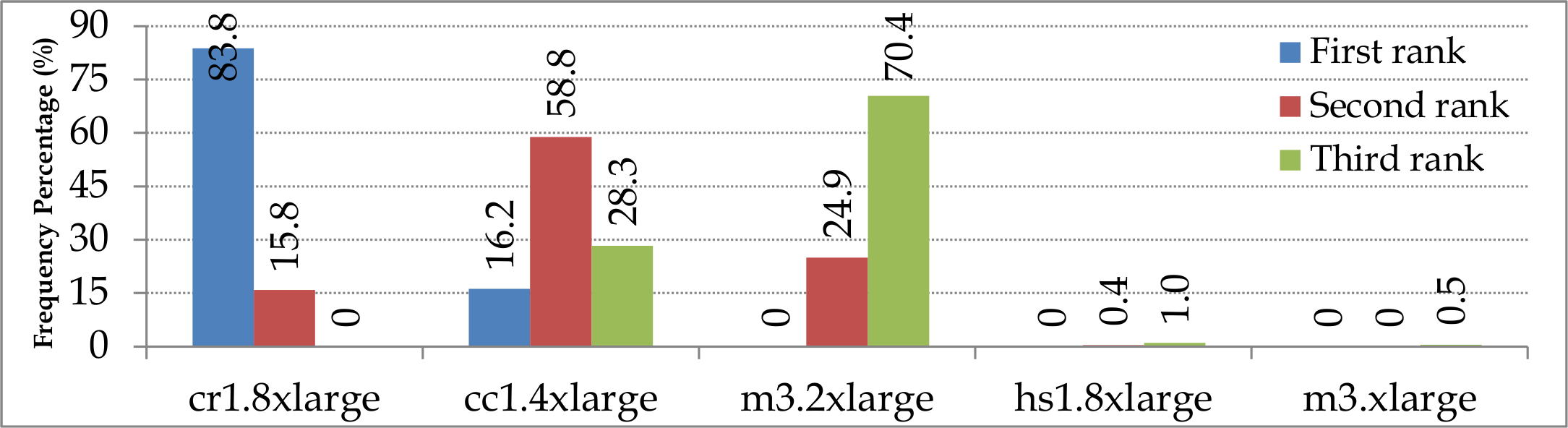}} \\
	\subfigure[P ranking: parallel execution]{\label{figure51a}\includegraphics[width=0.46\textwidth]{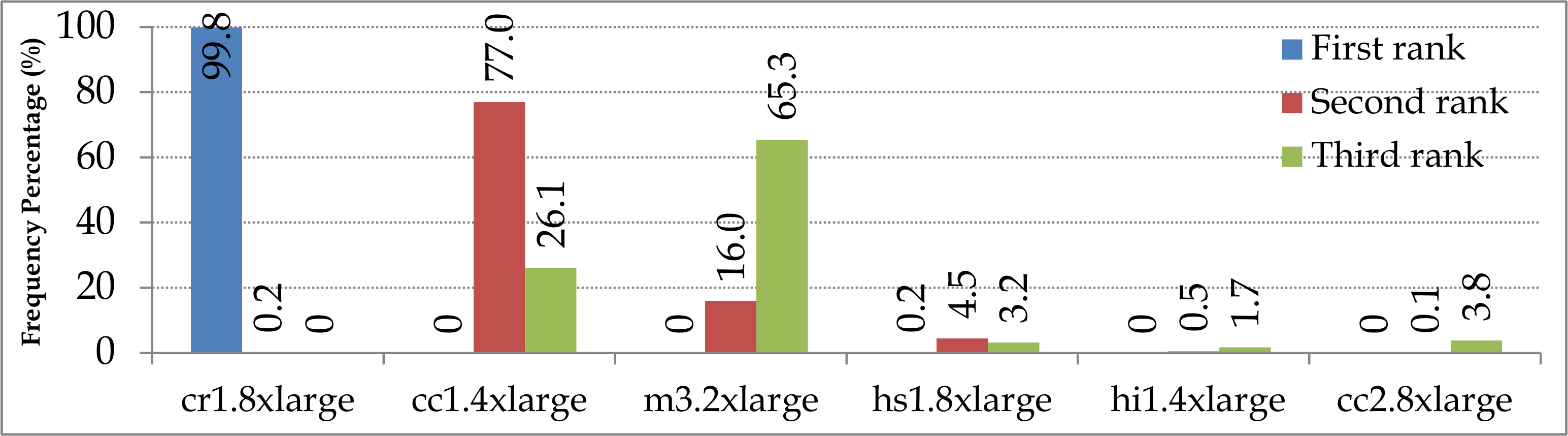}} \\
	\subfigure[PC ranking: sequential execution]{\label{figure50b}\includegraphics[width=0.46\textwidth]{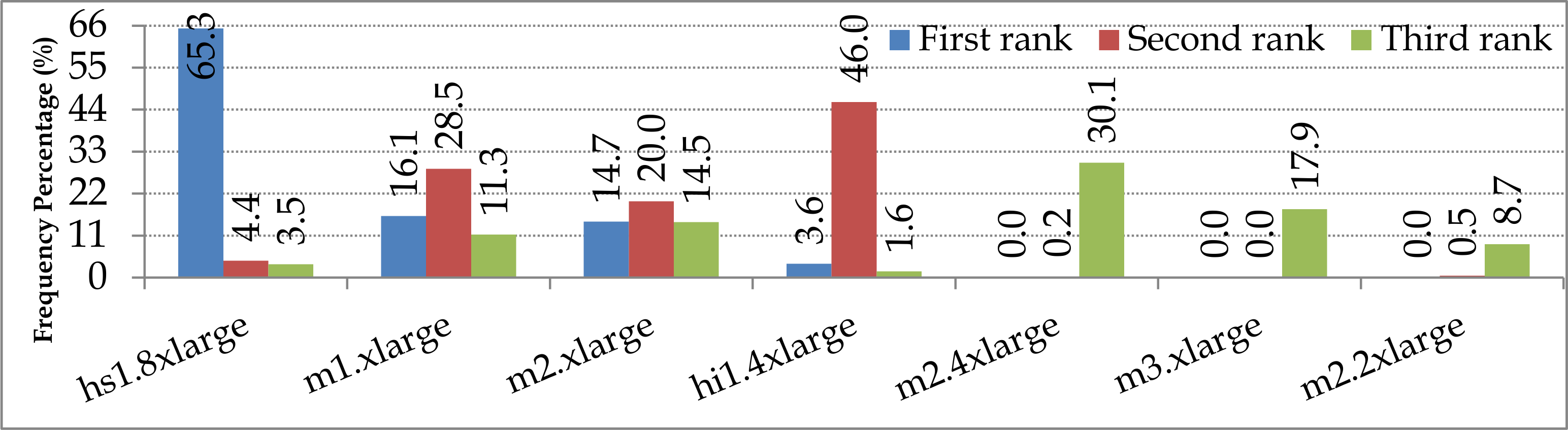}}\\
	\subfigure[PC ranking: parallel execution]{\label{figure51b}\includegraphics[width=0.46\textwidth]{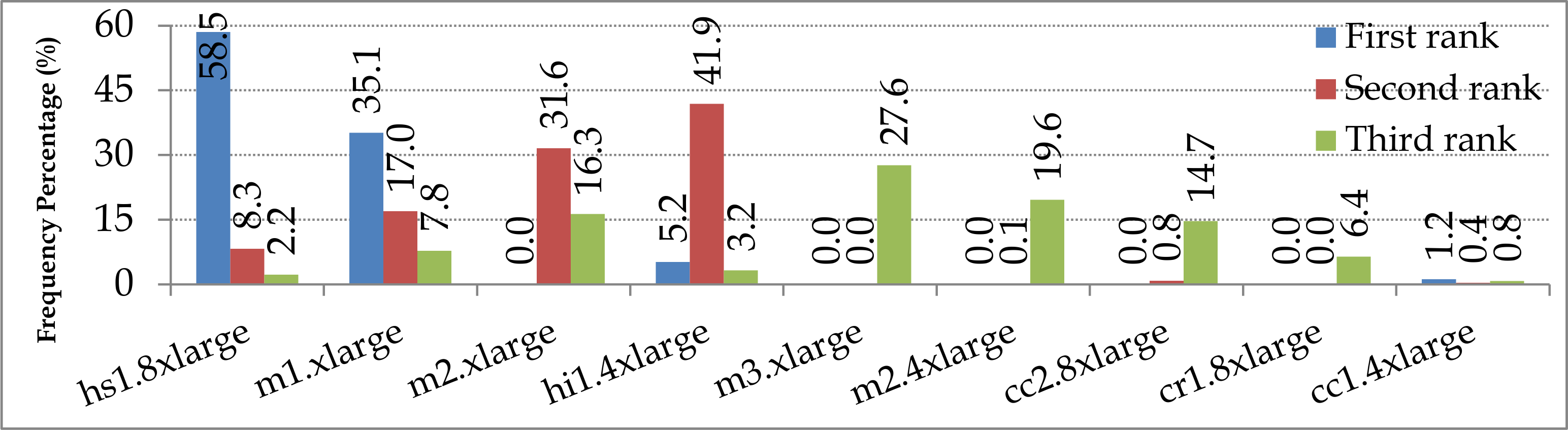}}
\caption{Frequency of instances appearing in the top 3 ranks for 1,295 combination of aggregate weights}
\label{figure50}
\end{figure}

\subsection{Aggregate Weight Space}
Figure \ref{figure50} shows the frequency of the instances appearing in the top three ranks in the aggregate weight space. When the P ranks are taken into account, it is evident that there are winners for the first, second and third ranks (refer Figure \ref{figure50a} and Figure \ref{figure51a}). This indicates that the  benchmarking methodology is not highly sensitive to small variations in the weight combinations. So given an application that a user has limited knowledge about, the methodology is likely to produce ranks that will provide maximum performance.

When cost is taken into account with performance (PC), there is a more uniform spread across the top three ranks (refer Figure \ref{figure50b} and Figure \ref{figure51b}). We observe the following: (a) four instances appear in the first and second ranks in contrast to the former case with only one or two instances in the first and second ranks, (b) at least two instances that achieve first ranks have a good chance of winning, (c) two instances, \texttt{hs1} and \texttt{hi1} have better chances of winning when cost is considered, and (d) the \texttt{cr1} instance which is a clear winner in the P ranking has only a small chance of winning when cost is taken into account. The chances of winning are shared between instances when cost is accounted for. This makes the winning instances less obvious intuitively and the weights are relatively more sensitive compared to the P ranks. 

\begin{figure}
\centering
	\subfigure[P ranking: sequential execution]{\label{figure60a}\includegraphics[width=0.46\textwidth]{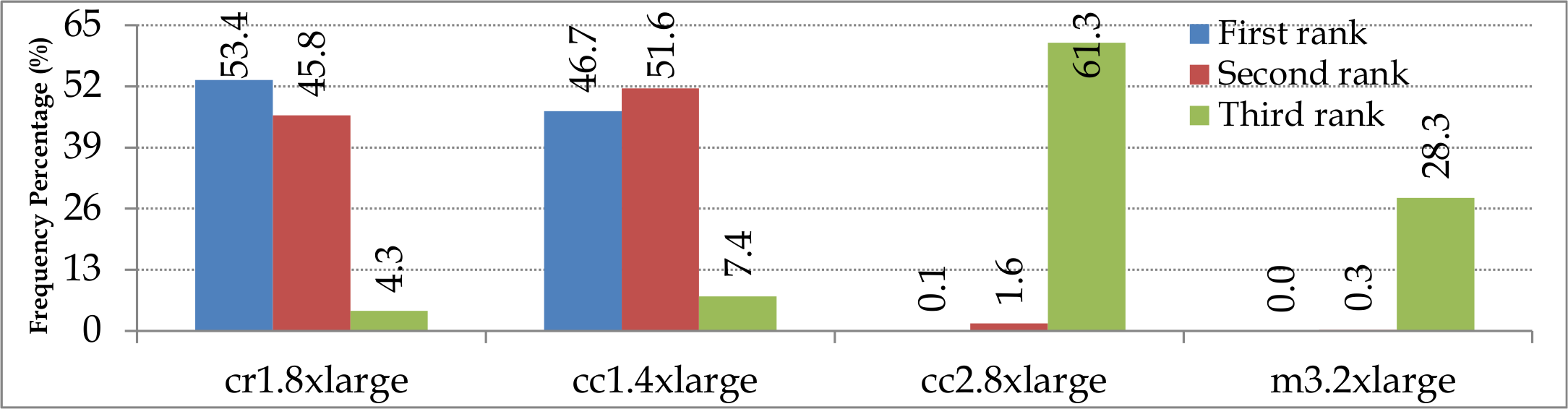}} \\
	\subfigure[P ranking: parallel execution]{\label{figure61a}\includegraphics[width=0.46\textwidth]{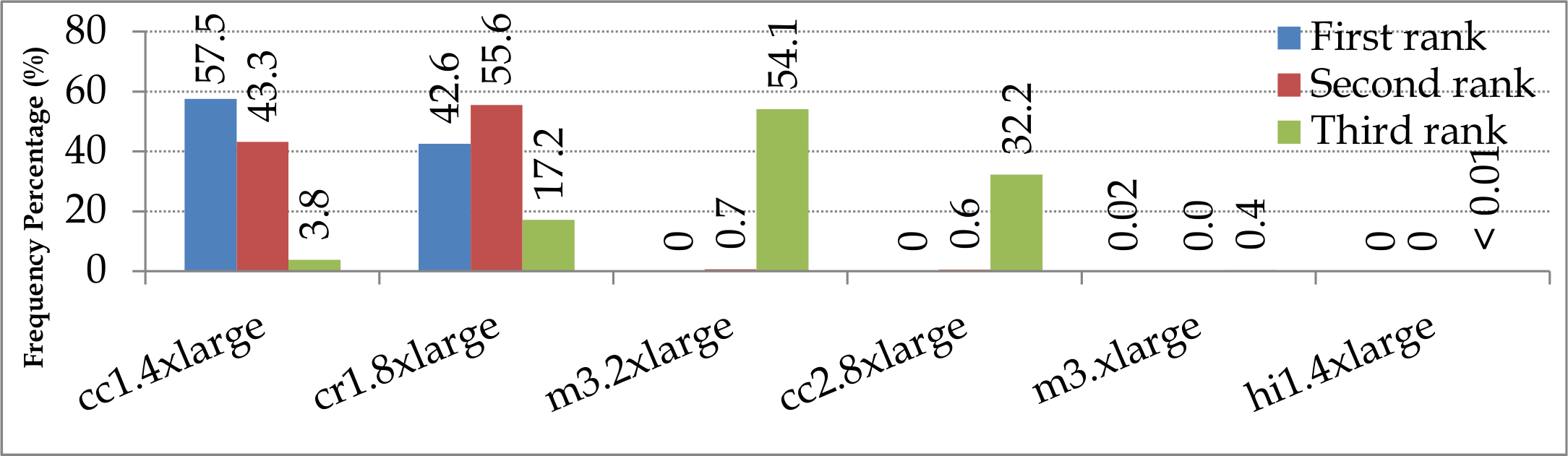}} \\
	\subfigure[PC ranking: sequential execution]{\label{figure60b}\includegraphics[width=0.46\textwidth]{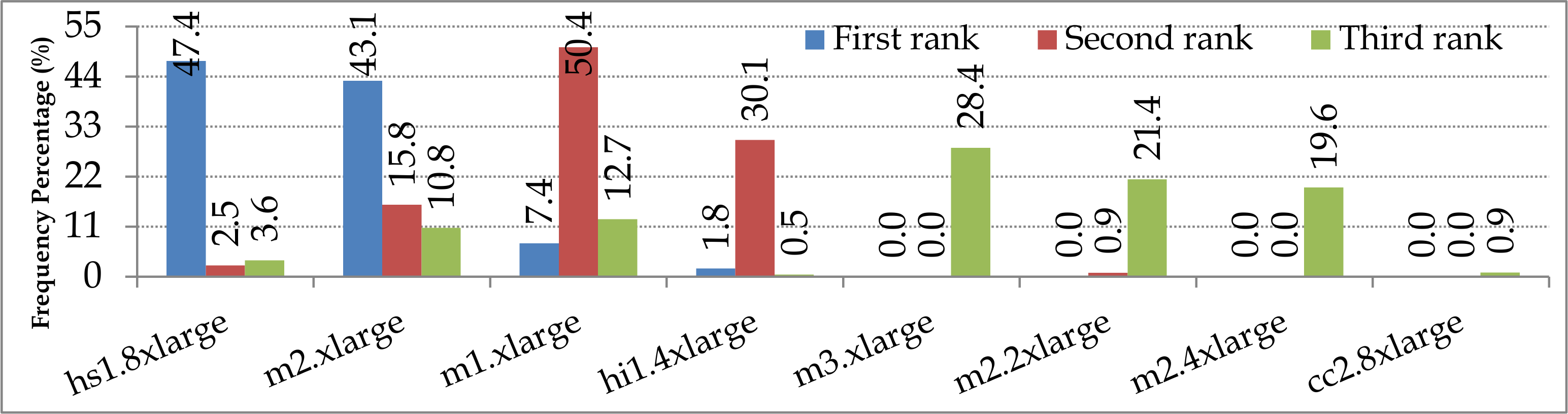}} \\
	\subfigure[PC ranking: parallel execution]{\label{figure61b}\includegraphics[width=0.46\textwidth]{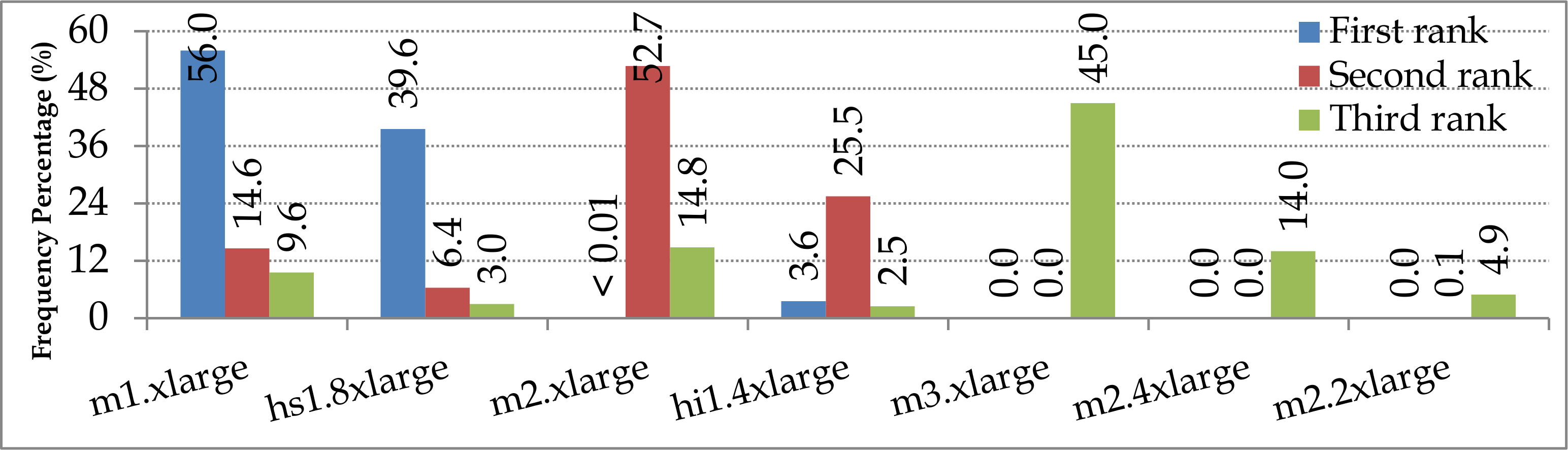}}
\caption{Frequency of instances appearing in the top 3 ranks for 1,679,615 combination of fine grain weights}
\label{figure60}
\end{figure}

\subsection{Fine Grain Weight Space}
Figure \ref{figure60} shows the frequency of the instances appearing in the top three ranks in the fine grain weight space. Unlike the aggregate weight space, there are more winners in this case. For example, the first and second ranks are almost equally shared between \texttt{cr1} and \texttt{cc1} instances and the third rank by \texttt{cg1} for sequential and parallel execution (refer Figure \ref{figure60a} and Figure \ref{figure61a}). As expected, the methodology produces a wider range of rankings when a larger number of groups and associated weights are considered. 

The number of competing instances increases when cost is taken into account (refer Figure \ref{figure60b} and Figure \ref{figure61b}). At least two instances have a good chance of winning the first rank and at least three instances occupy the second rank. The third rank instances generally do not win first or second positions. The key difference from the aggregate weight space is the wide spread of winning instances which can be selected by the benchmarking methodology. There is a greater sensitivity to the set of weights supplied by the user but at the same time facilitates the selection of instances, which are both performance and cost effective, for an application.

\section{Validation Study}
\label{sensitivityanalysis}  
In this section, we examine the benchmarking methodology for case study applications using validation techniques. 

\subsection{Case Studies}

The benchmarking methodology we have proposed is suitable to predict the performance of embarrassingly parallel high-performance computing applications that execute on the multiple cores of the same VM. Three such case studies are chosen for validating the benchmarking methodology which can be executed on multiple VMs. However, in this research, we execute the applications on all virtual cores available on a single VM. The first application is a simulation used in the financial risk industry, the second application is a molecular dynamics simulations used by theoretical physicists, and the third is a mathematical solver employed in scientific applications. The applications have different requirements and are therefore chosen for validating the methodology. 

\subsubsection{Case Study 1: Financial Risk Analysis}
Aggregate Risk Analysis \cite{agganalysis-1,agganalysis-2} is a simulation employed for computing key risk metrics such as Probable Maximum Loss (PML) \cite{pml-1}. The simulation is developed using C++ and parallelised on the CPU cores using OpenMP. The Boost library is used for implementing mathematical and statistical functions. The inputs to the simulation are one million catastrophic event trials and a collection of thousands of events and their corresponding losses which are obtained from the disk onto memory. A number of financial terms are applied to the loss associated with an event and aggregated to provide the PML for a contractual year. The simulation is embarrassingly parallel and can be executed on the virtual cores by sharing the input data between the threads on the core.  

The simulation is memory intensive with numerous read and write operations and at the same time computationally intensive requiring a large number of float operations to be performed both to compute the risk metrics. The local communication between processes are less relevant. The simulation requires data transfer from the disk to memory initially making file operations moderately relevant. 
 
\subsubsection{Case Study 2: Molecular Dynamic Simulation}
The second case study is a molecular dynamics simulation of short range interactions used by theoretical physicists of a complex system comprising $N$ particles \cite{md-1, md-2}. The simulation is developed using C++ and OpenMP is leveraged for parallelism on the cores of the CPU. The simulation computes the trajectory of $N$ particles and the forces they exert by solving a system of differential equations discretized into different time steps. In each time step, the position, velocity, the kinetic and potential energies of each particle are generated. It is assumed that if particles collide then they pass through each other. The simulations are performed on a three dimensional space for 10,000 particles and 200 time steps.

The simulation is computationally intensive followed by the memory and processor requirements along with the need for local communication. There are no file intensive operations in this simulation.

\subsubsection{Case Study 3: Block Triagonal Solver}
The Block Triagonal Solver, otherwise referred to as BT is a NASA Advanced Supercomputing (NAS) Parallel Benchmark (NPB) \cite{npb-1}, version 3.3.1\footnote{\url{https://www.nas.nasa.gov/publications/npb.html}}, which is used as a third case study. The BT solver is a pseudo application benchmark that represents a sample mathematical solver used in scientific applications. The Class C problem (defines the size and parameters of the problem) of the solver is used. Sequential and parallel programming models (in this paper, OpenMP is used) are used for an empirical analysis on the cloud VMs using NPB. The solver uses a grid size of $162 \times 162 \times 162$ for 200 iterations.

The solver is numerically intensive and memory and processor related operation is relevant, but does not take a precedence over computations. Local communications and file operations have little effect on the solver. 

\subsection{Validation Techniques}
\label{sec:validationtechniques}
Two validation techniques are used to evaluate the benchmarking methodology. The first technique compares the ranking produced by the methodology and the ranking generated from empirical analysis of the application, which we refer to as `Comparative Validation'. In the second technique, the entire space of weights is taken into account to make observations when comparing the aggregate and fine-grain weight spaces, which we refer to as `Enumeration-based Validation'.

\subsubsection{Comparative Validation}
This technique comprises the following three steps:

\textit{Step 1 - Application Benchmarking:} the case study applications are used for an empirical analysis to validate the cloud benchmarking methodology. The application is firstly executed on all VMs. The time taken by an application to complete execution can be used as one criterion for evaluating performance.

\textit{Step 2 - Application-based Cloud Ranking:} based on the performance of the VMs in the empirical analysis they are ranked. For example, in this paper, the ranks from the empirical analysis are based on performance evaluated in terms of the time taken for completing execution. Additional criterion such as the quality of result can be used if it is applicable to the application. 

The values for each criterion for evaluating performance are normalised using $\bar{v}_{i, j} = \frac{v_{i, j} - \mu_{j}}{\sigma_{j}}$, where $\mu_j$ is the mean value of $v_{i, j}$ over $m$ VMs and $\sigma_j$ is the standard deviation $v_{i, j}$ over $m$ VMs. 
The normalised values are used to rank the VMs $Mp_{i}$.
If multiple criteria are used then a rank for each criterion is obtained. The multiple ranks then need to be aggregated into a single rank.

\textit{Step 3 - Cloud Ranks Comparison:} the ranks $R_{i}$ from the benchmarking methodology are compared against the validation ranks $M_{i}$. This comparison can be used for selecting one or more cloud resources that can maximise the performance of the application. If there are significant differences between the rankings of the VMs then the requirements of the application are re-evaluated and a different set of weights need to be assigned to the attribute groups.

\begin{figure*}
\centering
	\subfigure[Case study 1: sequential]{\label{figure4-1}\includegraphics[width=0.325\textwidth]{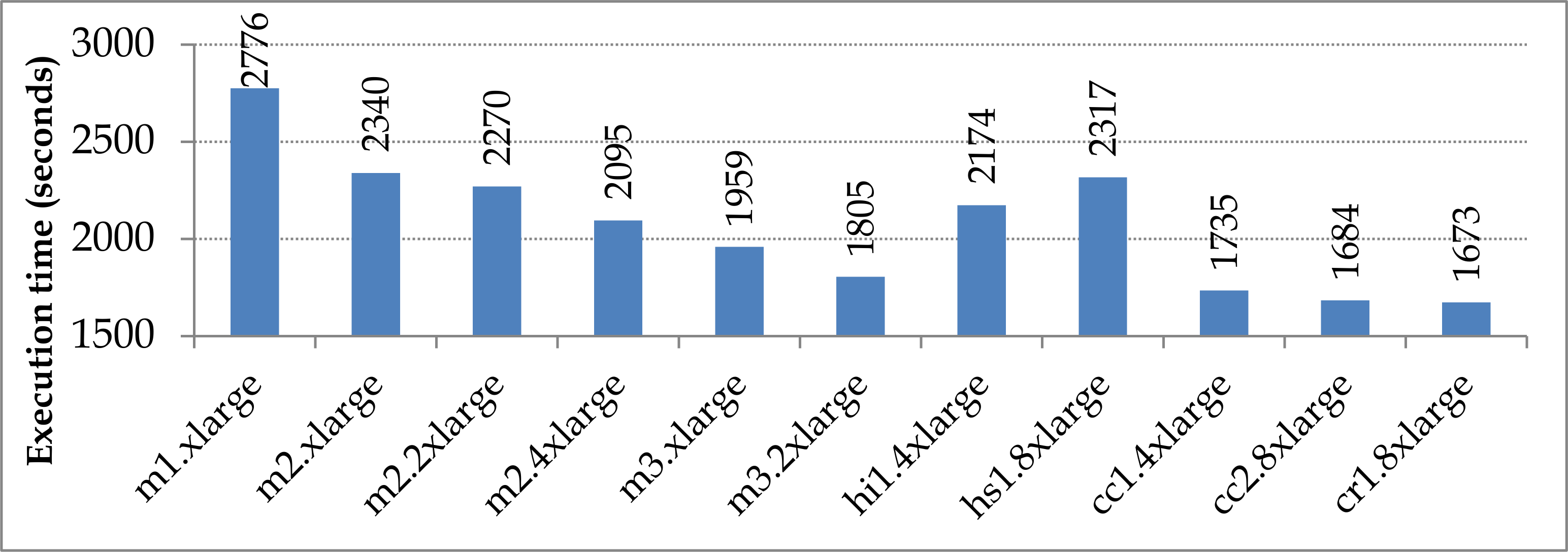}} \hfill
	\subfigure[Case study 2: sequential]{\label{figure5-1}\includegraphics[width=0.325\textwidth]{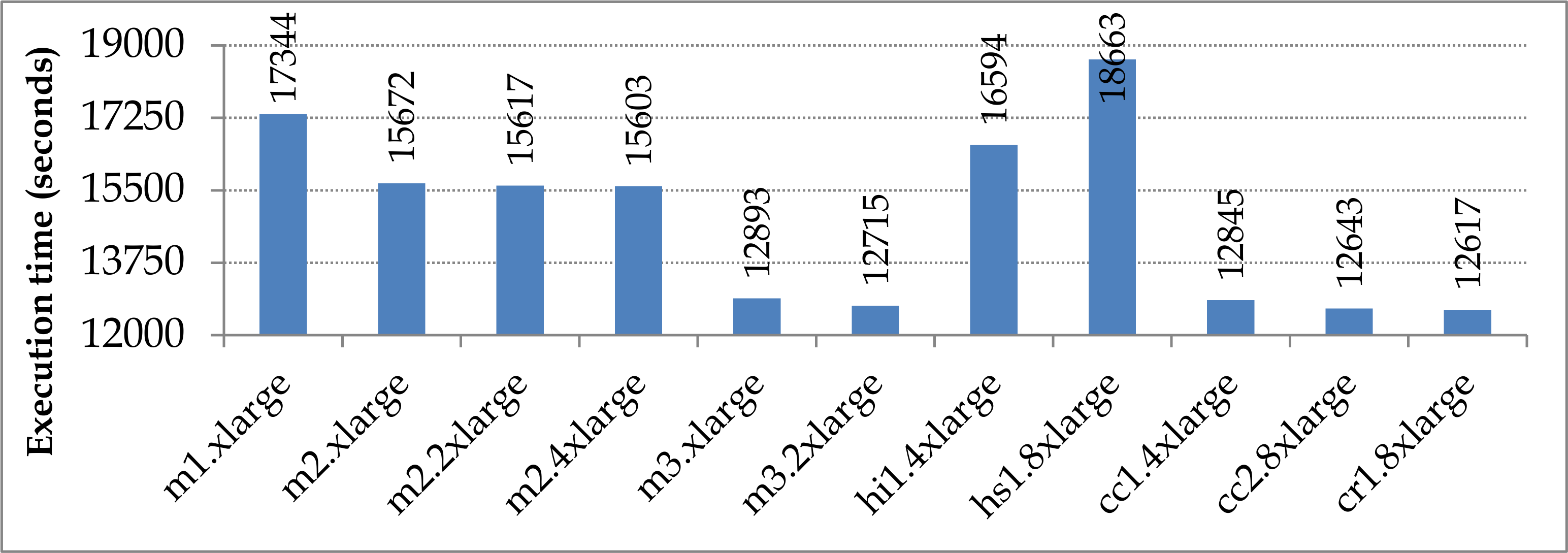}} \hfill
	\subfigure[Case study 3: sequential]{\label{figure6-1}\includegraphics[width=0.325\textwidth]{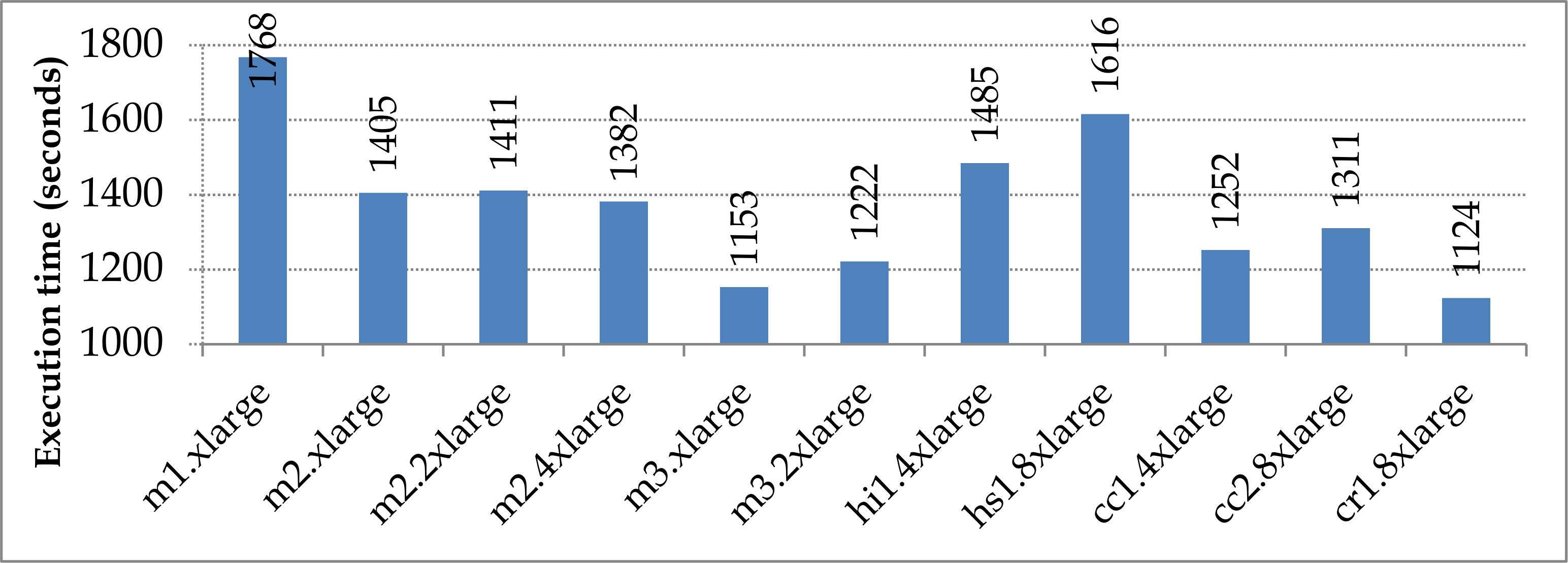}}\\
	\subfigure[Case study 1: parallel]{\label{figure4-2}\includegraphics[width=0.325\textwidth]{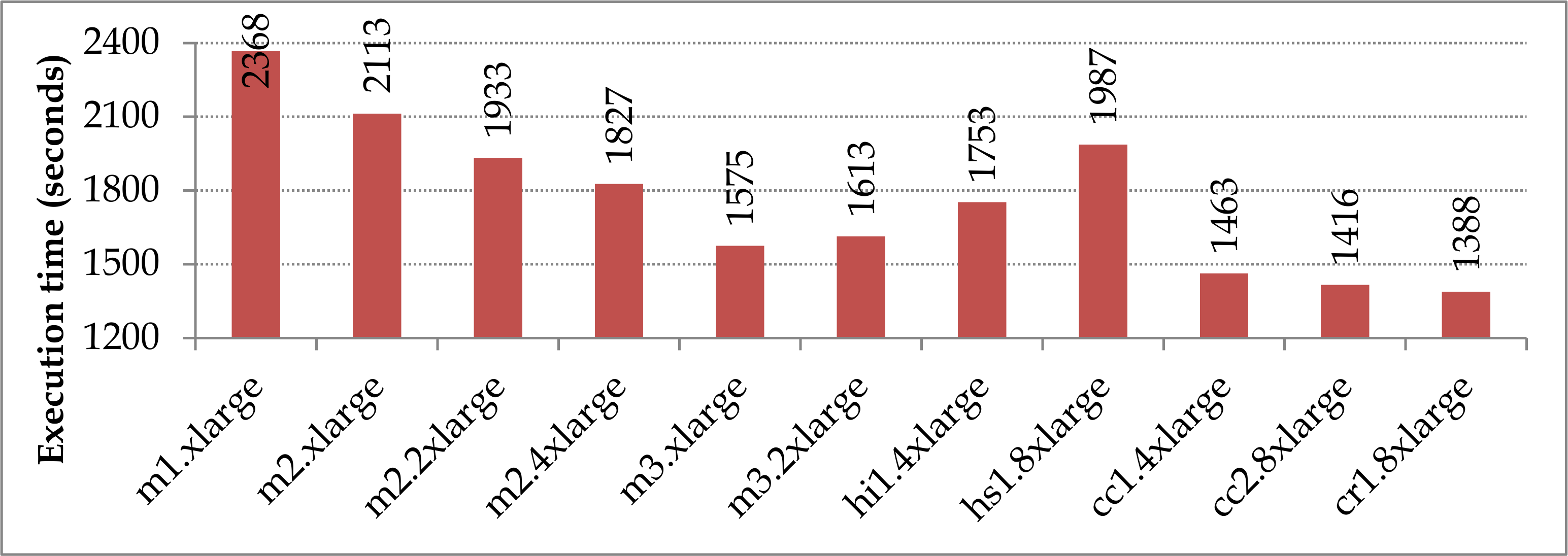}} \hfill
	\subfigure[Case study 2: parallel]{\label{figure5-2}\includegraphics[width=0.325\textwidth]{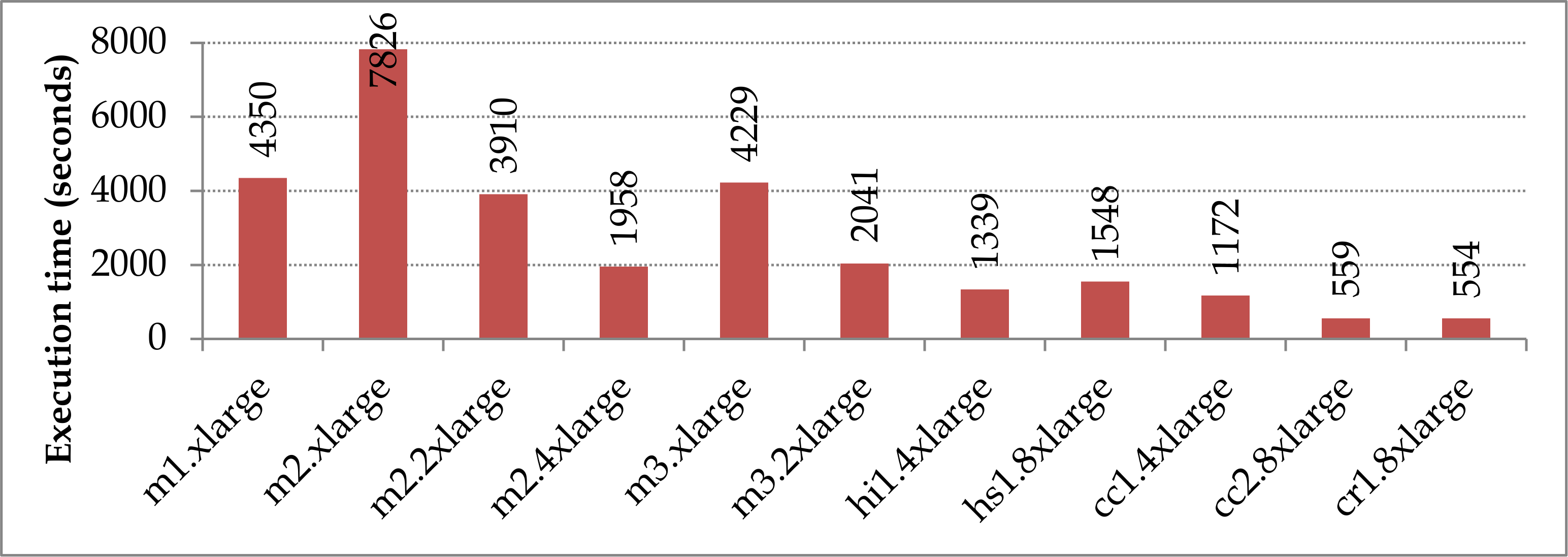}} \hfill
	\subfigure[Case study 3: parallel]{\label{figure6-2}\includegraphics[width=0.325\textwidth]{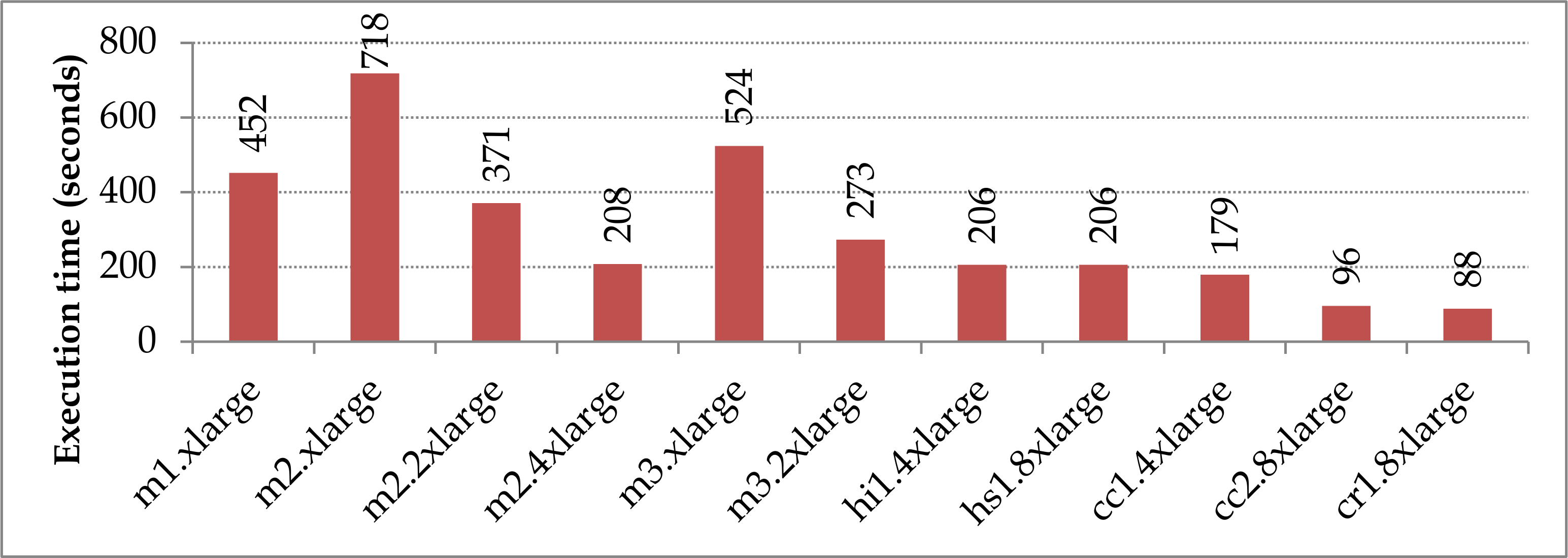}}
\caption{Time taken for sequential and parallel execution (min 2 threads and max 32 threads) of the case studies}
\label{figure456}
\end{figure*}


The three case study applications were executed on all instances shown in Table \ref{table1}. The mean time taken to execute the applications ten times were obtained. Figure \ref{figure456} shows the time taken for the sequential execution and parallel execution (the case study applications are executed on all the cores of the VM; 2 cores minimum and 32 cores maximum) of the applications respectively. For the P rankings, the instances are ranked based on their execution time; first rank for instance with lowest execution time. For the PC rankings, the ratio of the cost per hour of an instance (from Table \ref{table1}) and the execution time (in this case is the measure of performance) is ordered in ascending order to generate the ranks. Again, the first rank is for the instance with the lowest ratio. The rankings of the methodology for each case study given a set of weights chosen in consultation with industry practitioners and domain experts were also generated. 
In this paper, the results using fine-grain rankings are considered. Aggregate weight-based ranking has been previously reported and $W_{agg}=\{5, 3, 5, 2\}$ for case study 1, $W_{agg}=\{4, 3, 5, 0\}$ for case study 2 and $W_{agg}=\{2, 0, 5, 0\}$ for case study 3 \cite{cloudbenchmark-0}. 


Table \ref{comparativetablecasestudy1} shows the rankings for the financial risk application with $W_{sub}=\{3, 5, 3, 3, 2, 5, 2, 2\}$. There is a correlation of nearly 86\% and 57\% for sequential and parallel P ranks respectively. For the top four VMs (sequential performance) the benchmarking methodology points to the same instances observed as good performers in the empirical analysis. In the case of parallel performance we observe that the top four VMs are similar in both the benchmarking methodology and the empirical analysis. When costs are taken into account the correlation is improved to 93\% and 91\%.


Table \ref{comparativetablecasestudy2} shows the rankings for the molecular dynamics simulation with $W_{sub}=\{3, 4, 3, 3, 2, 5, 0, 0\}$. There is a correlation of nearly 85\% and over 67\% for sequential performance and parallel P ranks respectively. 
For parallel performance the top five instances in the empirical analysis are pointed to by the benchmarking methodology. 


Table \ref{comparativetablecasestudy3} shows the rankings for the BT solver with $W_{sub}=\{2, 2, 0, 0, 5, 5, 0, 0\}$. 
There is 81\% and 95\% correlation for sequential performance. For parallel performance the rankings bear 95\% correlation. 

The percentage correlation (obtained by using the Pearson Product-Moment Correlation Coefficient) between the ranks obtained from the benchmarking methodology and the empirical analysis are shown in Table \ref{correlationtable2}. There is high correlation between the ranks, which is an indication that the benchmarking methodology with fine-grain weights can produce P and PC ranks that are close to reality as verified through the case studies. The robustness of the methodology is confirmed through this validation exercise.

\begin{table*}
	\centering
	\caption{Fine-grain Rankings for Case Study 1, $W_{sub}=\{3, 5, 3, 3, 2, 5, 2, 2\}$}
	\begin{tabular}{| p{2.2cm} | p{1.3cm} | p{1.3cm} | p{1.3cm} | p{1.3cm} | p{1.3cm} | p{1.4cm} | p{1.3cm} | p{1.3cm} |}
		\hline
		\multirow{2}{*}{Amazon Instance}	&	\multicolumn{2}{c |}{Sequential P Rankings}	& \multicolumn{2}{c |}{Parallel P Rankings} &	\multicolumn{2}{c |}{Sequential PC Rankings}	& \multicolumn{2}{c |}{Parallel PC Rankings}\\
		\cline{2-9}
		&	From Benchmarking	&	From Empirical Analysis	&	From Benchmarking 	&	From Empirical Analysis &	From Benchmarking	&	From Empirical Analysis	&	From Benchmarking &	From Empirical Analysis\\
		\hline
		\hline
		m1.xlarge	&	10	&	11	&	11	&	11	&	2	&	1	&	1	&	2	\\
		m2.xlarge	&	9	&	10	&	9	&	10	&	1	&	2	&	2	&	1	\\
		m2.2xlarge	&	8	&	8	&	8	&	8	&	3	&	4	&	3	&	5	\\
		m2.4xlarge	&	7	&	6	&	7	&	7	&	9	&	7	&	5	&	6	\\
		m3.xlarge	&	5	&	5	&	6	&	4	&	4	&	3	&	4	&	3	\\
		m3.2xlarge	&	3	&	4	&	3	&	5	&	5	&	5	&	6	&	4	\\
		hi1.4xlarge	&	6	&	7	&	4	&	6	&	8	&	8	&	9	&	8	\\
		hs1.8xlarge	&	11	&	9	&	10	&	9	&	10	&	11	&	11	&	11	\\
		cc1.4xlarge	&	2	&	3	&	1	&	3	&	6	&	6	&	8	&	7	\\
		cc2.8xlarge	&	4	&	2	&	5	&	2	&	7	&	9	&	10	&	10	\\
		cr1.8xlarge	&	1	&	1	&	2	&	1	&	11	&	10	&	7	&	9	\\

		
		\hline
	\end{tabular}
	\label{comparativetablecasestudy1}
\end{table*}

\begin{table*}
	\centering
	\caption{Fine-grain Rankings for Case Study 2, $W_{sub}=\{3, 4, 3, 3, 2, 5, 0, 0\}$}
	\begin{tabular}{| p{2.2cm} | p{1.3cm} | p{1.3cm} | p{1.3cm} | p{1.3cm} | p{1.3cm} | p{1.4cm} | p{1.3cm} | p{1.3cm} |}
		\hline
		\multirow{2}{*}{Amazon Instance}	&	\multicolumn{2}{c |}{Sequential P Rankings}	& \multicolumn{2}{c |}{Parallel P Rankings} &	\multicolumn{2}{c |}{Sequential PC Rankings}	& \multicolumn{2}{c |}{Parallel PC Rankings}\\
		\cline{2-9}
		&	From Benchmarking &	From Empirical Analysis	&	From Benchmarking &	From Empirical Analysis &	From Benchmarking &	From Empirical Analysis	&	From Benchmarking &	From Empirical Analysis\\
		\hline
		\hline
		m1.xlarge	&	11	&	10	&	10	&	10	&	1	&	2	&	2	&	2	\\
		m2.xlarge	&	7	&	8	&	9	&	11	&	2	&	1	&	1	&	1	\\
		m2.2xlarge	&	9	&	7	&	8	&	8	&	4	&	4	&	4	&	4	\\
		m2.4xlarge	&	8	&	6	&	6	&	6	&	7	&	7	&	6	&	6	\\
		m3.xlarge	&	4	&	5	&	4	&	9	&	3	&	3	&	3	&	3	\\
		m3.2xlarge	&	3	&	3	&	7	&	7	&	5	&	5	&	5	&	5	\\
		hi1.4xlarge	&	6	&	9	&	3	&	4	&	10	&	9	&	10	&	8	\\
		hs1.8xlarge	&	10	&	11	&	11	&	5	&	8	&	10	&	8	&	9	\\
		cc1.4xlarge	&	2	&	4	&	5	&	3	&	6	&	6	&	7	&	7	\\
		cc2.8xlarge	&	5	&	2	&	1	&	2	&	9	&	8	&	9	&	10	\\
		cr1.8xlarge	&	1	&	1	&	2	&	1	&	11	&	11	&	11	&	11	\\
		\hline
	\end{tabular}
	\label{comparativetablecasestudy2}
\end{table*}

\begin{table*}
	\centering
	\caption{Fine-grain Rankings for Case Study 3, $W_{sub}=\{2, 2, 0, 0, 5, 5, 0, 0\}$}
	\begin{tabular}{| p{2.2cm} | p{1.3cm} | p{1.3cm} | p{1.3cm} | p{1.3cm} | p{1.3cm} | p{1.4cm} | p{1.3cm} | p{1.3cm} |}
		\hline
		\multirow{2}{*}{Amazon Instance}	&	\multicolumn{2}{c |}{Sequential P Rankings}	& \multicolumn{2}{c |}{Parallel P Rankings} &	\multicolumn{2}{c |}{Sequential PC Rankings}	& \multicolumn{2}{c |}{Parallel PC Rankings}\\
		\cline{2-9}
		&	From Benchmarking &	From Empirical Analysis	&	From Benchmarking &	From Empirical Analysis &	From Benchmarking &	From Empirical Analysis	&	From Benchmarking &	From Empirical Analysis\\
		\hline
		\hline
		m1.xlarge	&	11	&	11	&	11	&	11	&	1	&	1	&	2	&	3	\\
		m2.xlarge	&	7	&	8	&	8	&	10	&	2	&	2	&	1	&	1	\\
		m2.2xlarge	&	9	&	7	&	9	&	9	&	5	&	4	&	4	&	4	\\
		m2.4xlarge	&	8	&	6	&	7	&	7	&	7	&	7	&	7	&	7	\\
		m3.xlarge	&	5	&	2	&	5	&	6	&	4	&	3	&	3	&	2	\\
		m3.2xlarge	&	4	&	3	&	4	&	4	&	6	&	5	&	5	&	5	\\
		hi1.4xlarge	&	6	&	9	&	6	&	5	&	9	&	9	&	10	&	8	\\
		hs1.8xlarge	&	10	&	10	&	10	&	8	&	10	&	10	&	9	&	9	\\
		cc1.4xlarge	&	1	&	4	&	1	&	2	&	3	&	6	&	6	&	6	\\
		cc2.8xlarge	&	3	&	5	&	3	&	3	&	8	&	8	&	8	&	10	\\
		cr1.8xlarge	&	2	&	1	&	2	&	1	&	11	&	11	&	11	&	11	\\		
		\hline
	\end{tabular}
	\label{comparativetablecasestudy3}
\end{table*}

In the comparative validation method, the rankings calculated from the expert set of weights were compared against the empirical ranks. 
This method is limited for comparing the PC rankings in that two variables, namely time and cost, are used for obtaining the ranks. If any one of the variable has the same value, then the combination of performance-cost is not well captured. Additional variables will need to be captured to address this such that each VM is uniquely represented for a fairer benchmarking method than the method used to generate PC ranking in this paper. Nevertheless, additional variables will result in a more lengthy benchmarking process.   

\subsubsection{Enumeration-based Validation}
Comparative validation cannot explain the distribution of rankings in the entire space of different sets of weights. Therefore, we present an exhaustive enumeration method that takes the entire space of weights into account. 

The space of $1,295$ sets of aggregate weights and $1,679,615$ sets of fine grain weights are considered. The corresponding ranking of each set of weights is generated using the benchmarking methodology. A scoring mechanism is used for each calculated rank against the empirical ranks. A weighted hamming distance is used for ranking in Algorithm \ref{algorithm1}.

\begin{algorithm} 
	\caption{Weighted hamming distance scores}
	\label{algorithm1}
	\begin{algorithmic}[1]
		\Procedure{Distance\textendash Scores}{$ER$, $CR$, $m$}{}
			\State score = 0
			\For{each $i$ $R \in ER$ }
				\State $C = m - R + 1$
				\State $R' = rankOf(CR, i)$
				\State $score += C * |R - R'|$
		\EndFor
	\EndProcedure
	\end{algorithmic}
\end{algorithm}

The input to the algorithm are two rankings - $ER$, which is the empirical rank and $CR$, which is the calculated rank and $m$, the number of cloud VMs. The distance between the empirical and calculated rank is computed as the sum of point-wise distances.

For example, consider that there are 11 VMs to be ranked and assume the empirical rank of one VM to be 4 and the calculated rank of the same VM to be 2, then the distance between them is 2. The coefficient $C$ is $11 - 4 + 1 = 8$. The contribution to the $score$ is $8 * 2 = 16$. Assume the empirical rank of another VM to be 10 and the calculated rank to be 8, now the distance is again 2. However, the coefficient is significantly lower, $C = 11 - 10 + 1 = 2$ and its contribution to the score is 4. Using such a mechanism we ensure that VMs ranked closer to the top in the empirical ranking are given higher weights in comparison to those ranked lower. The weight decay is linear and varies between 1 and the number of VMs considered, in our case $m=11$.
Smaller scores indicate that the calculated rank is closer to the empirical rank.

\begin{figure*} 
\begin{center}
	\subfigure[Sequential P]{\includegraphics[width=0.238\textwidth]{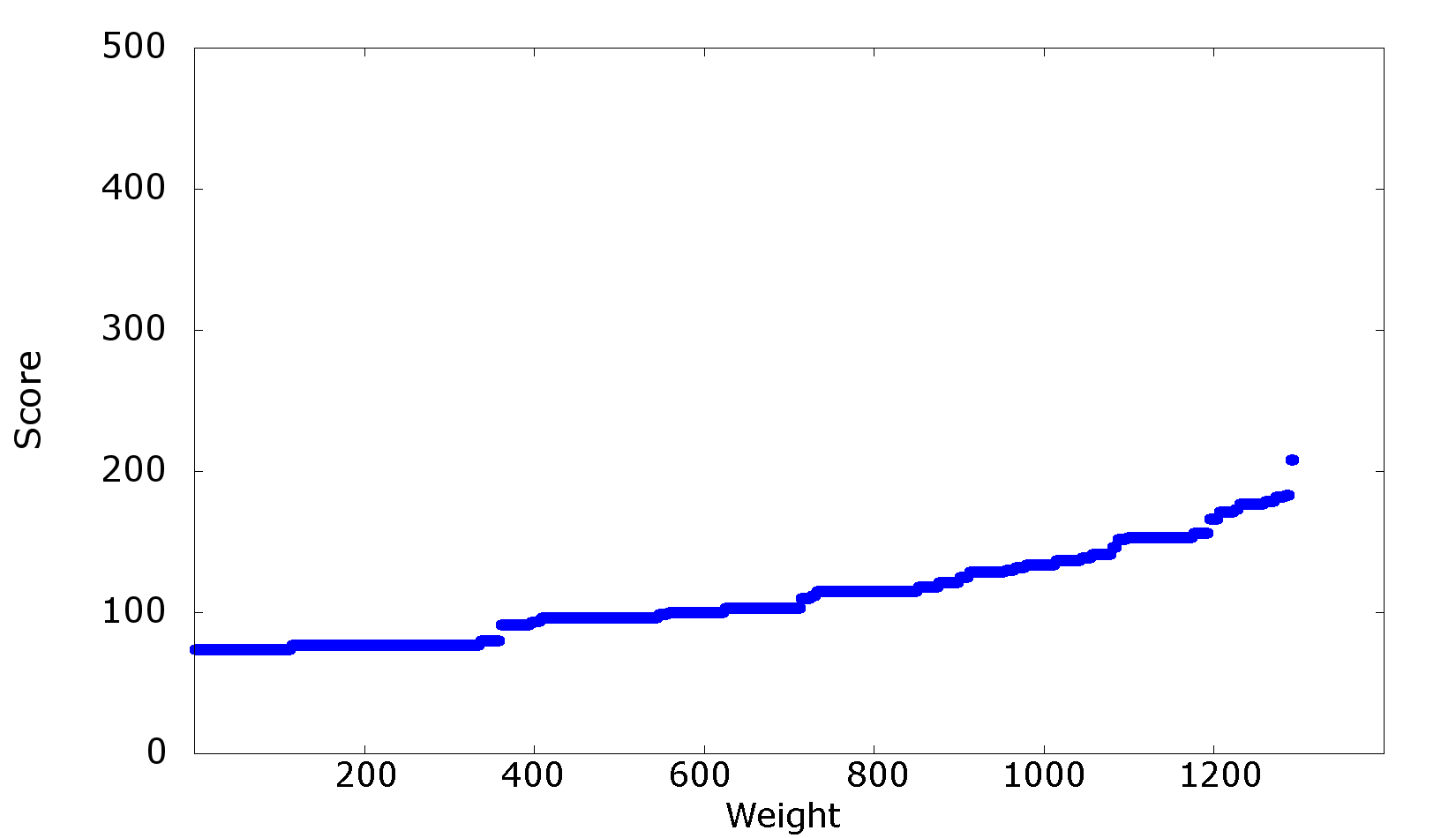} \label{figurec300-1}} \hfill
	\subfigure[Parallel P]{\includegraphics[width=0.238\textwidth]{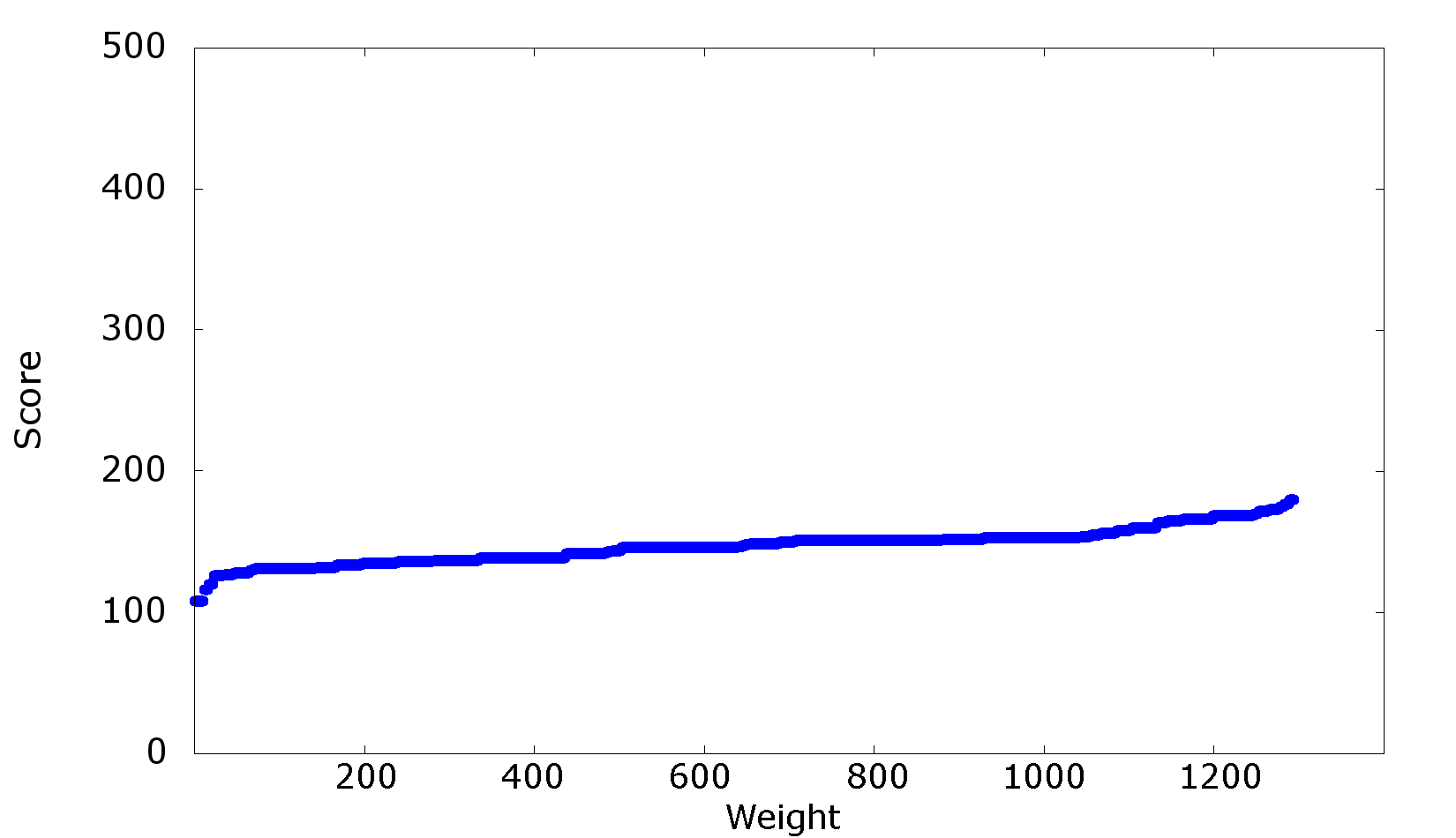} \label{figurec300-2}} \hfill
	\subfigure[Sequential PC]{\includegraphics[width=0.238\textwidth]{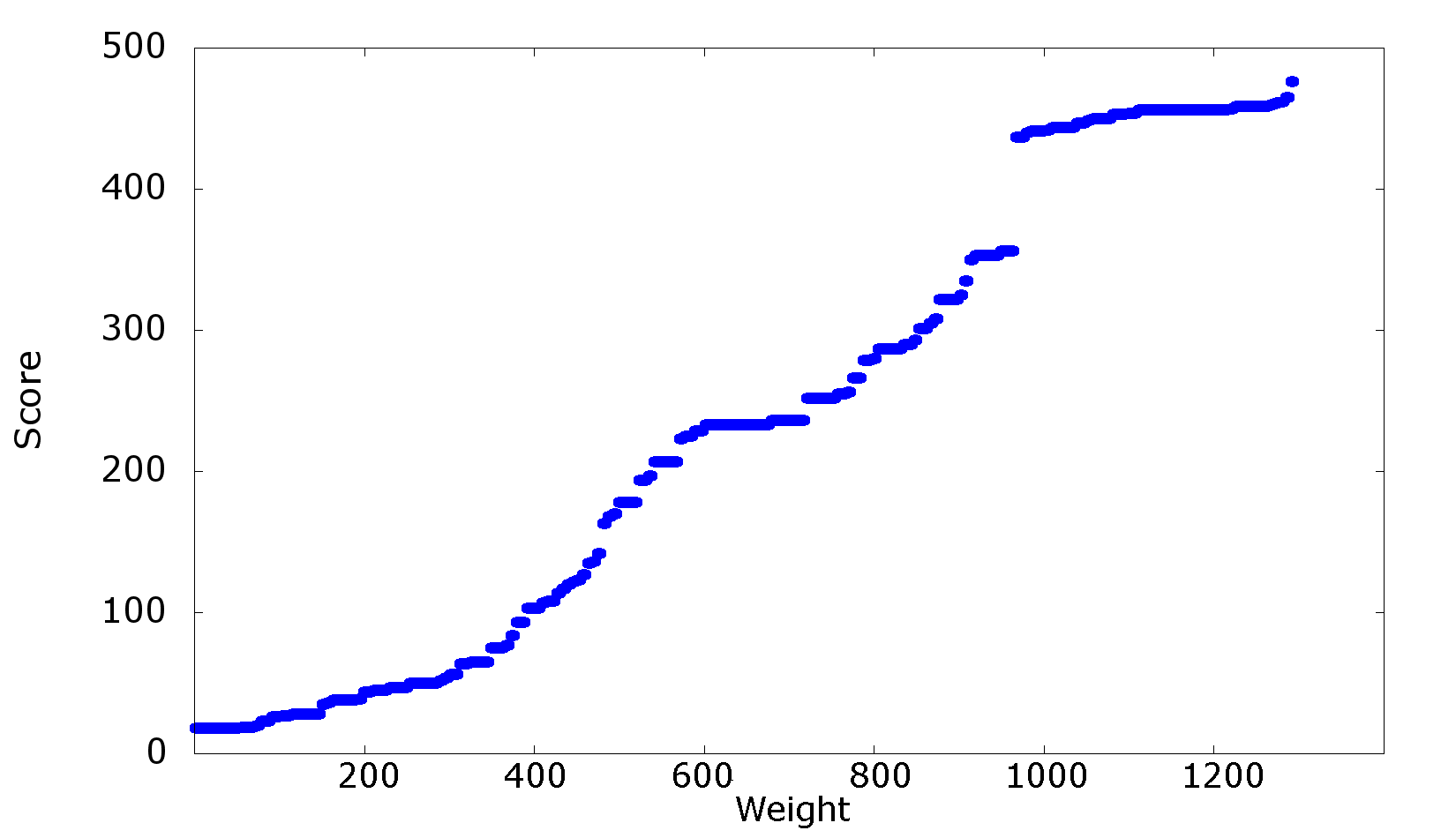} \label{figurec300-3}} \hfill
	\subfigure[Parallel PC]{\includegraphics[width=0.238\textwidth]{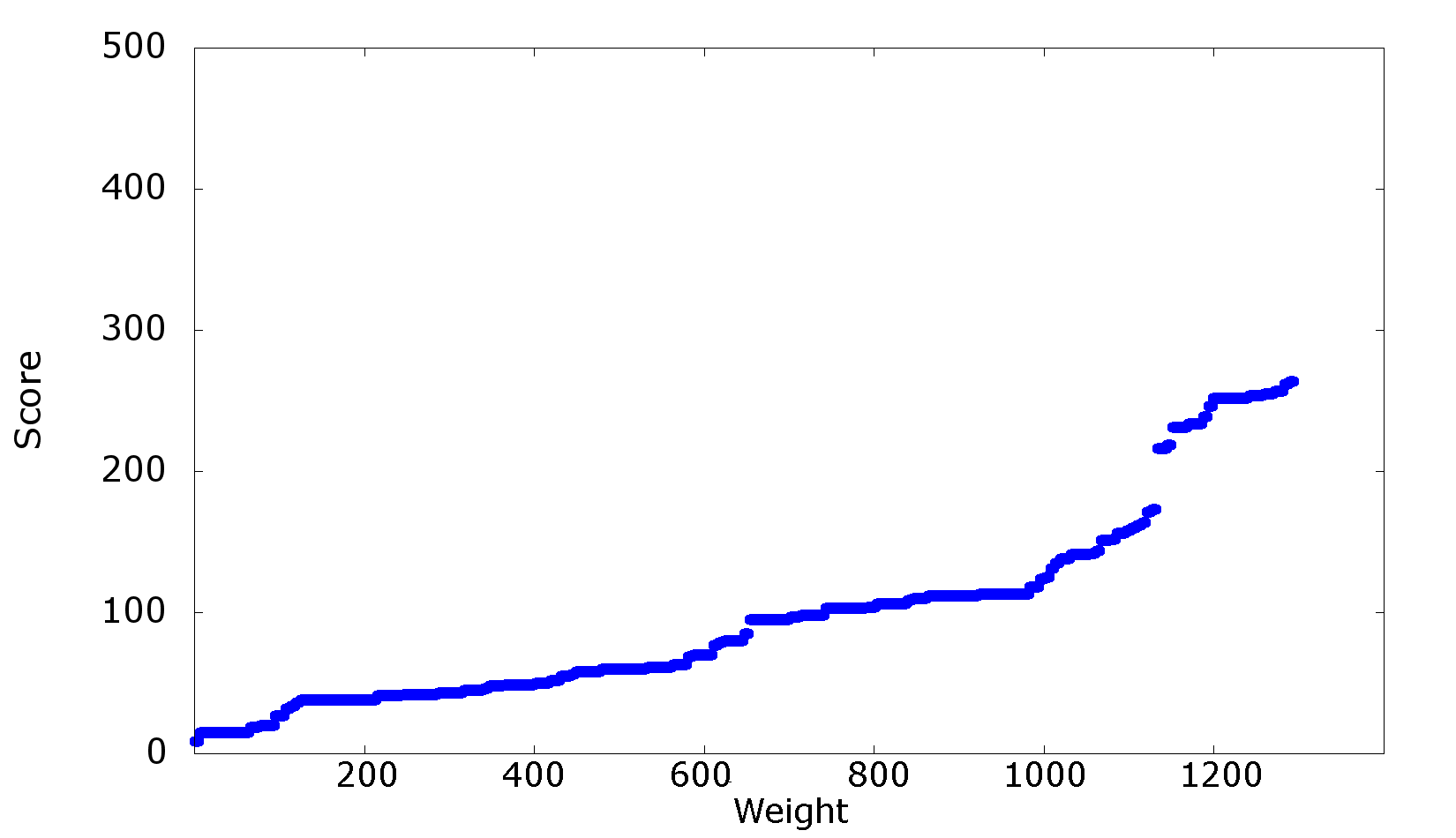} \label{figurec300-4}} \\
	\subfigure[Sequential P]{\includegraphics[width=0.238\textwidth]{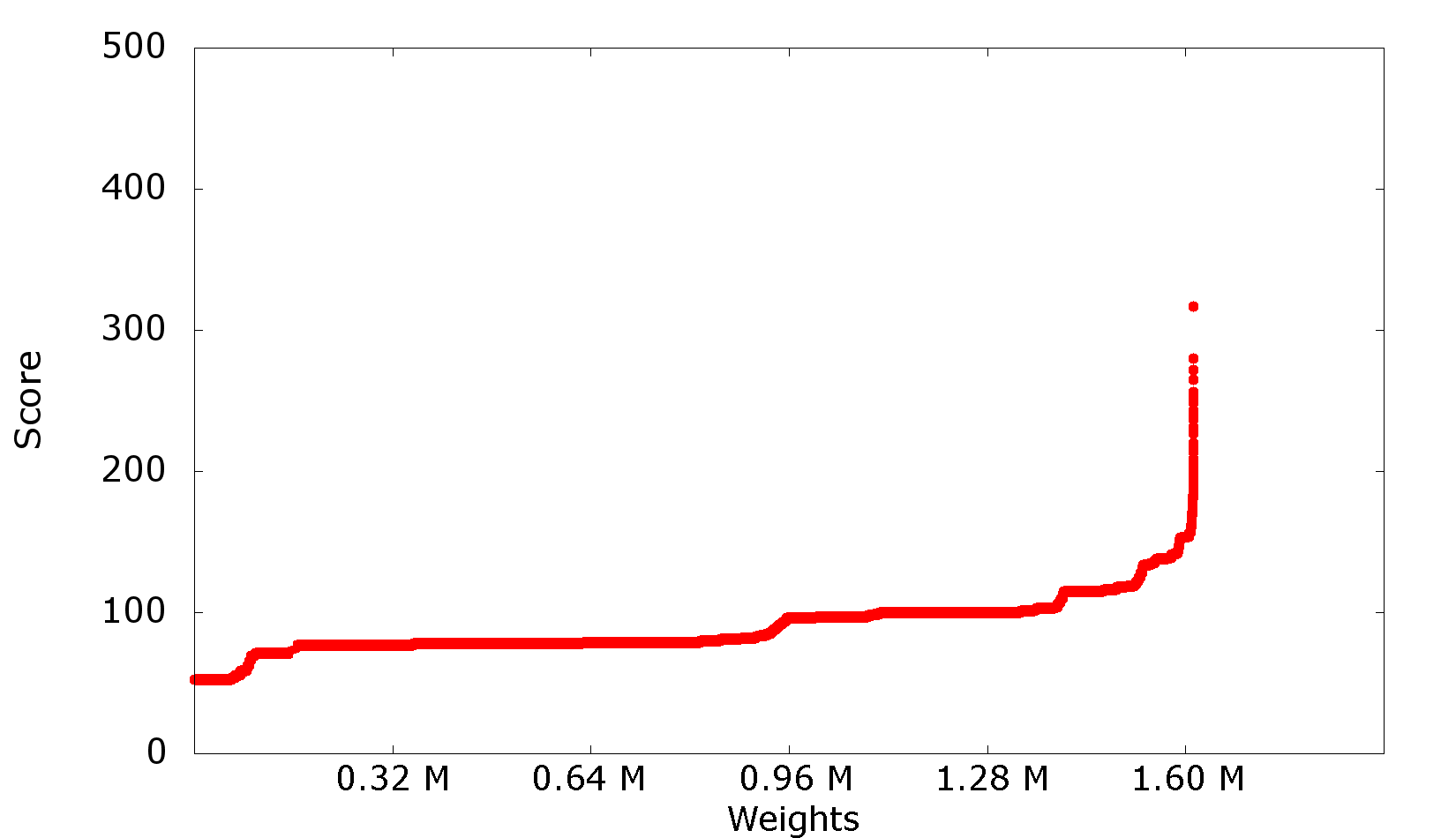} \label{figurec300-5}} \hfill
	\subfigure[Parallel P]{\includegraphics[width=0.238\textwidth]{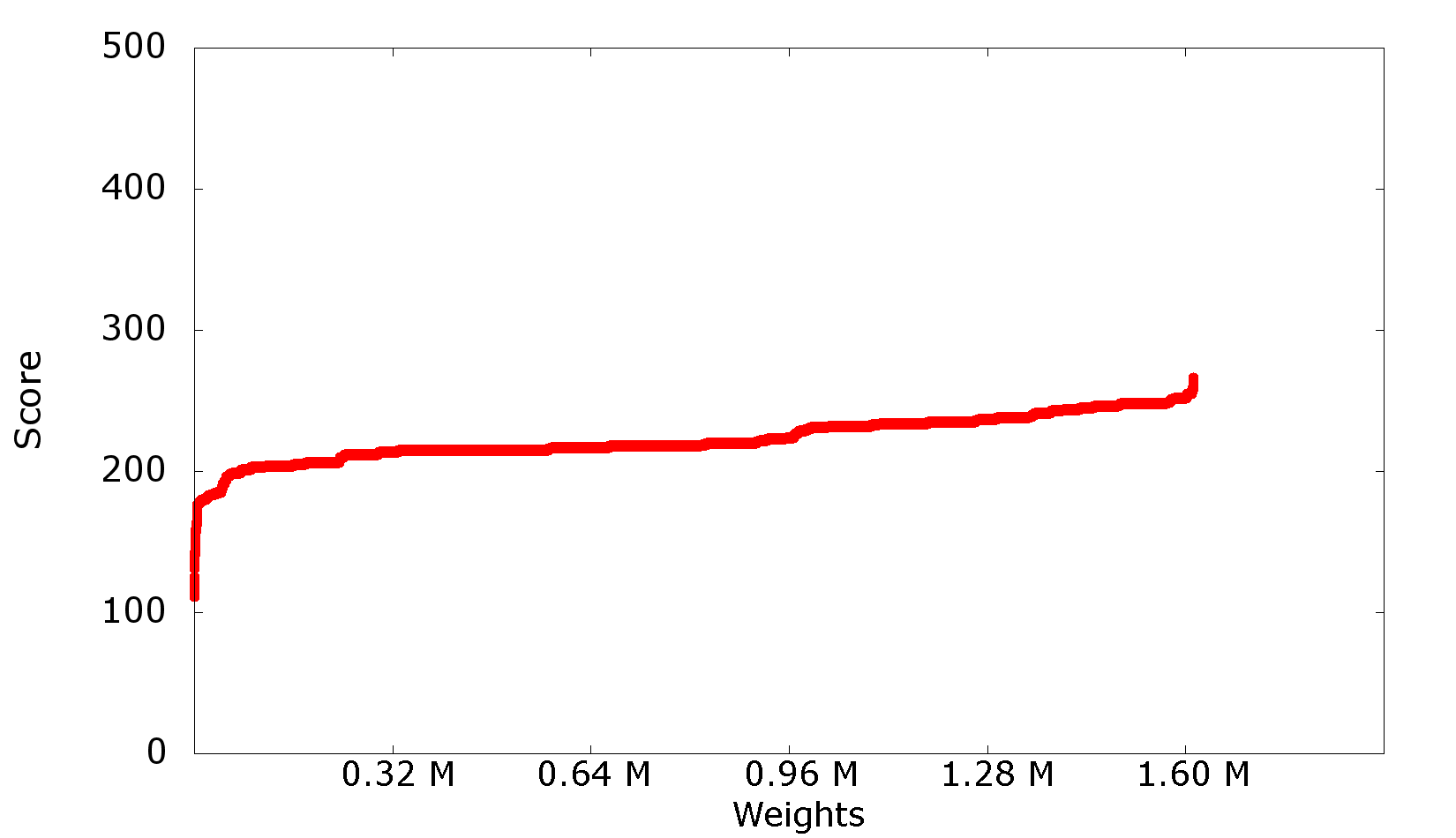} \label{figurec300-6}} \hfill
	\subfigure[Sequential PC]{\includegraphics[width=0.238\textwidth]{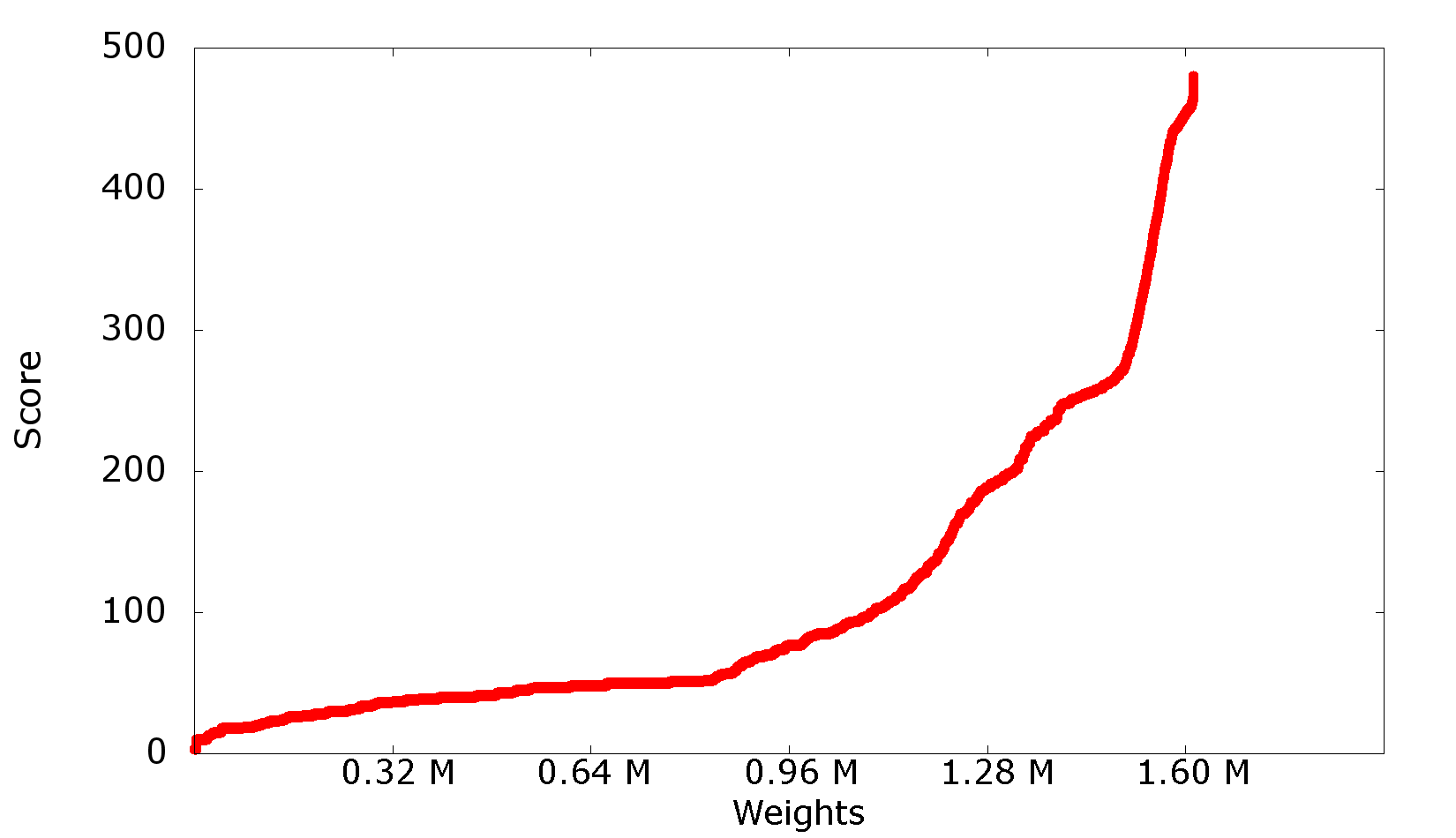} \label{figurec300-7}} \hfill
	\subfigure[Parallel PC]{\includegraphics[width=0.238\textwidth]{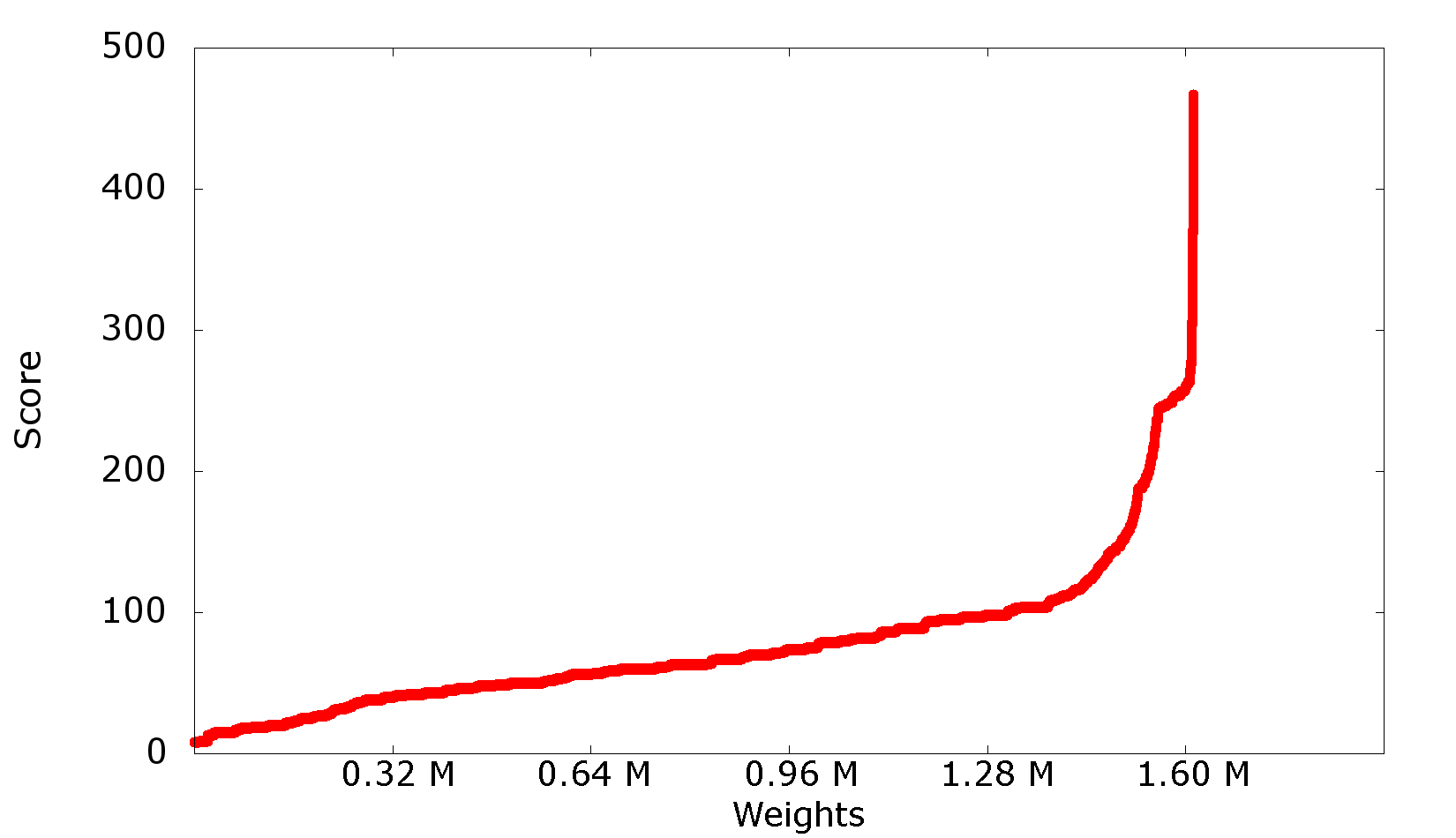} \label{figurec300-8}} \\
	\caption{Case Study 1 - Aggregate weights are shown in Figure \ref{figurec300-1} to Figure \ref{figurec300-4} and fine-grain weights are shown in Figure \ref{figurec300-5} to Figure \ref{figurec300-8}. X-axis is the weights ordered by scores shown in the Y-axis.}
	\label{figurec300}
\end{center}
\end{figure*}

\begin{figure*} 
\begin{center}
	\subfigure[Sequential P]{\includegraphics[width=0.238\textwidth]{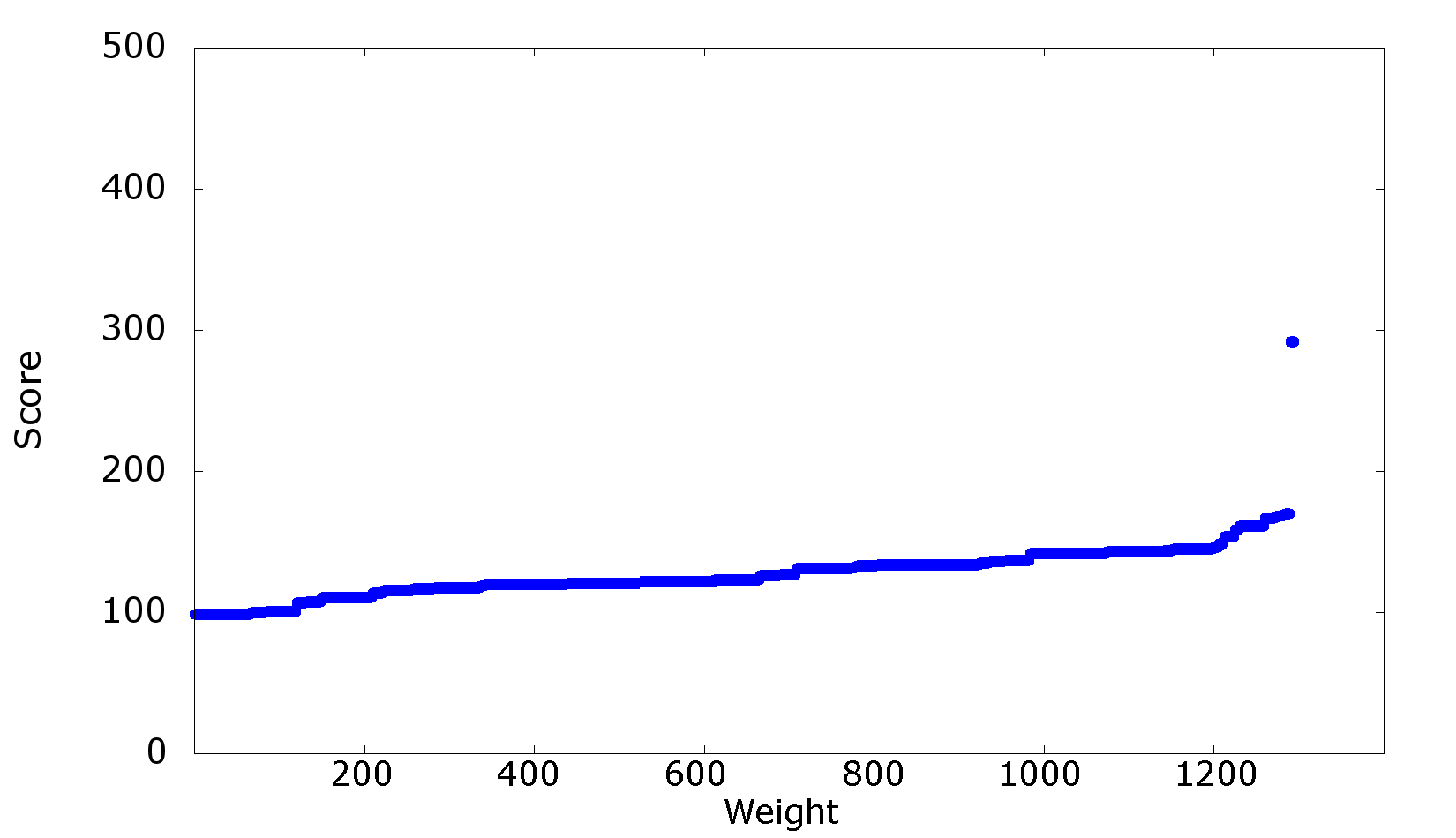} \label{figurec400-1}} \hfill
	\subfigure[Parallel P]{\includegraphics[width=0.238\textwidth]{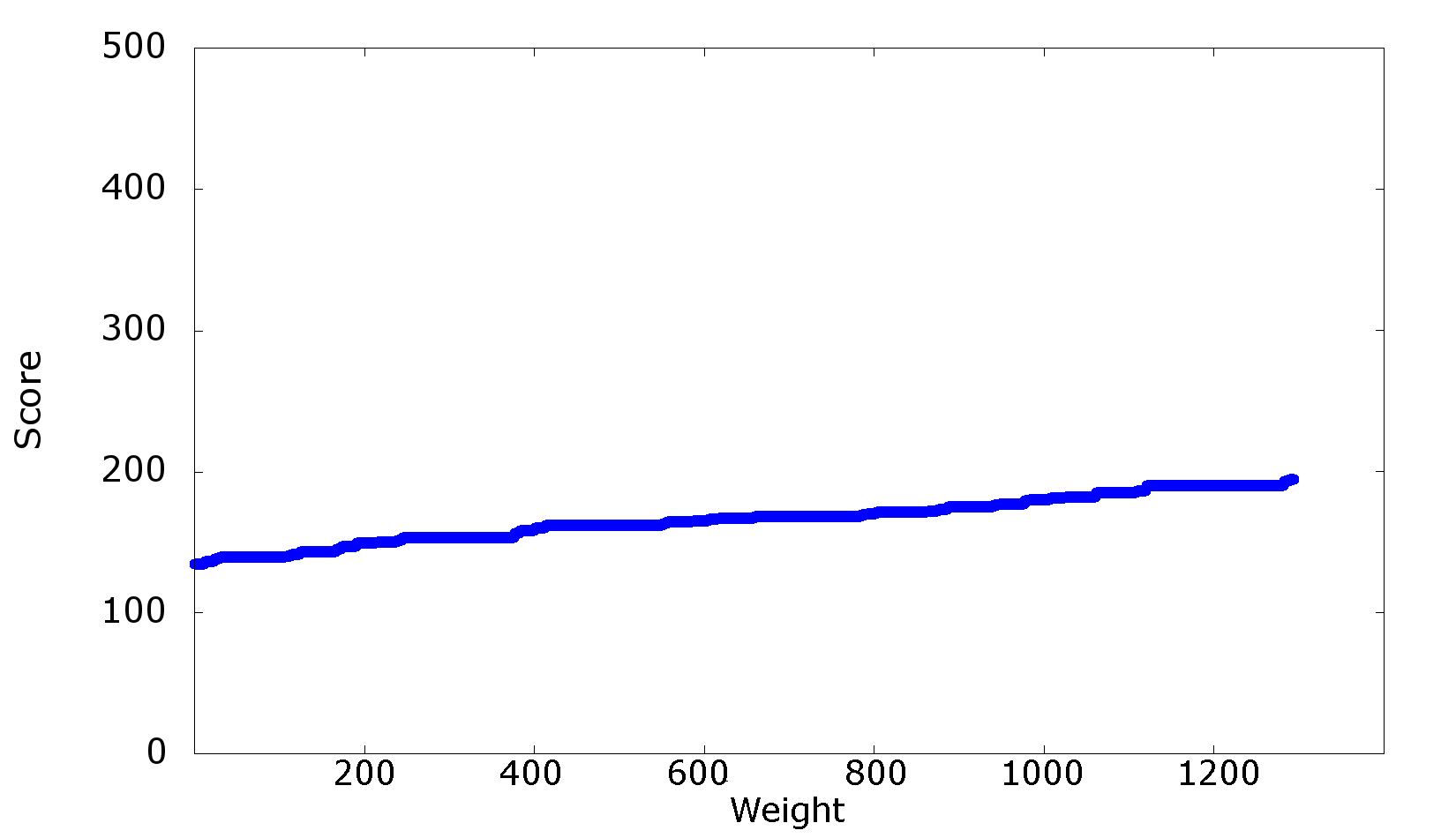} \label{figurec400-2}} \hfill
	\subfigure[Sequential PC]{\includegraphics[width=0.238\textwidth]{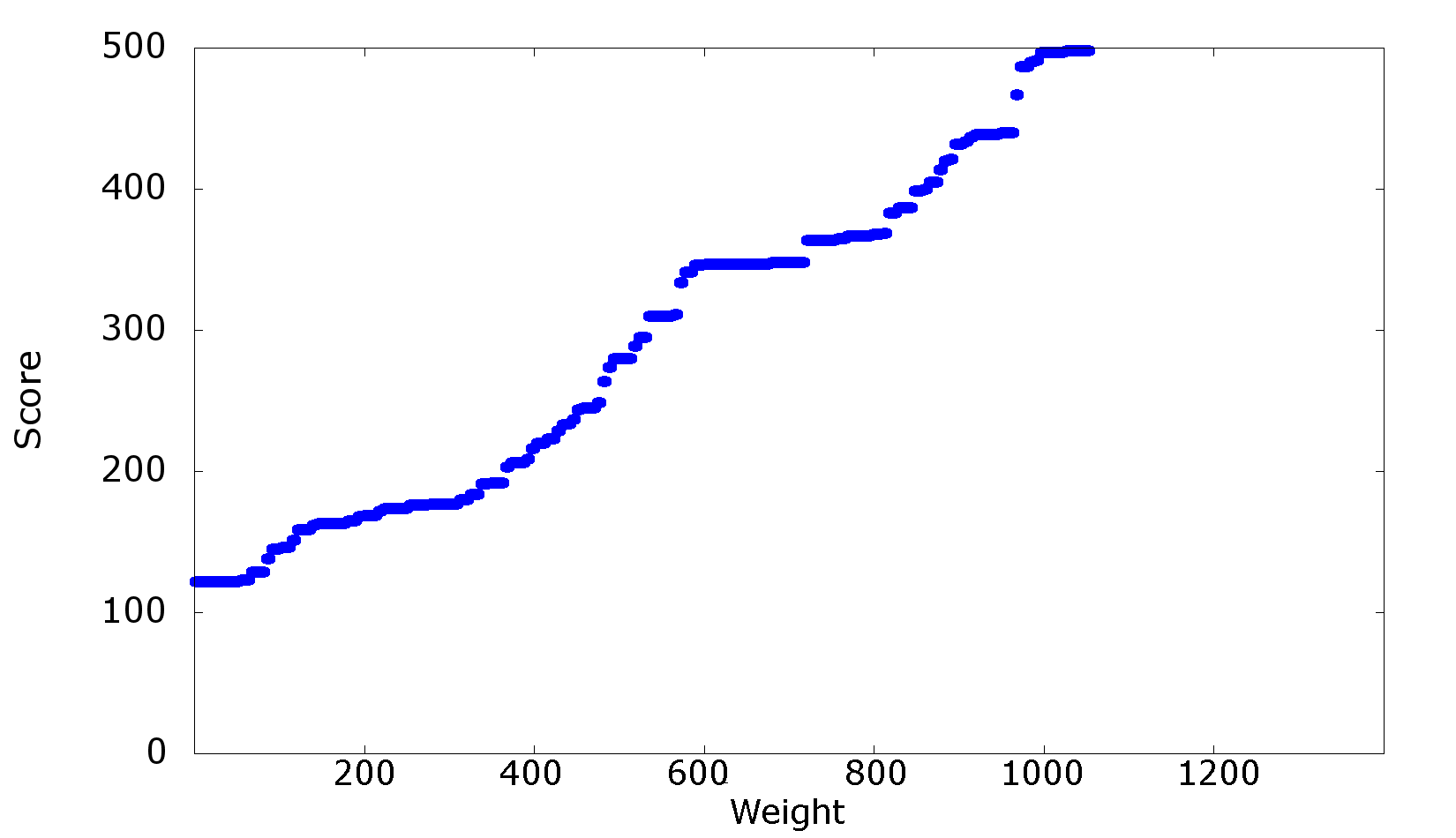} \label{figurec400-3}} \hfill
	\subfigure[Parallel PC]{\includegraphics[width=0.238\textwidth]{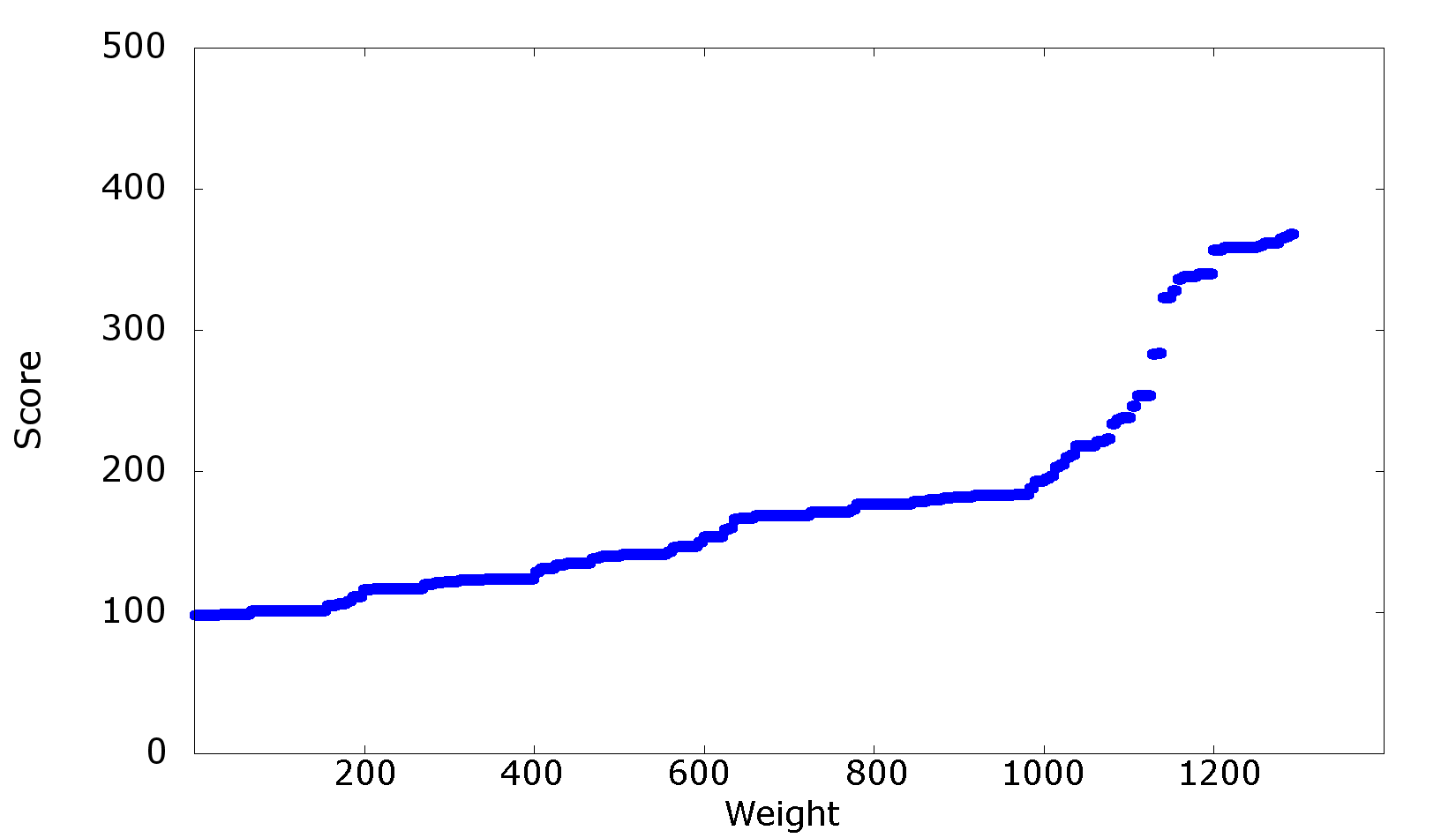} \label{figurec400-4}} \\
	\subfigure[Sequential P]{\includegraphics[width=0.238\textwidth]{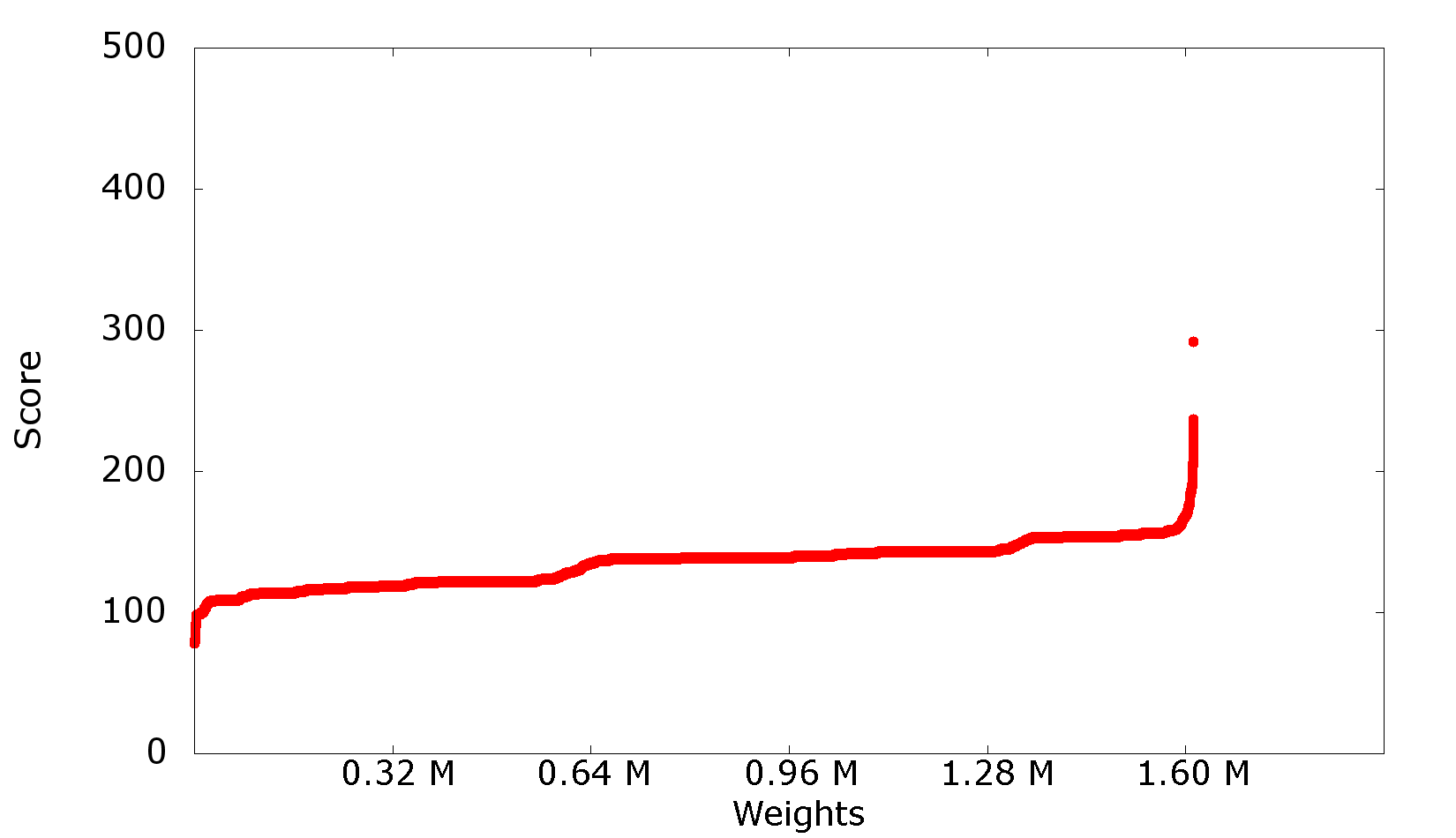} \label{figurec400-5}} \hfill
	\subfigure[Parallel P]{\includegraphics[width=0.238\textwidth]{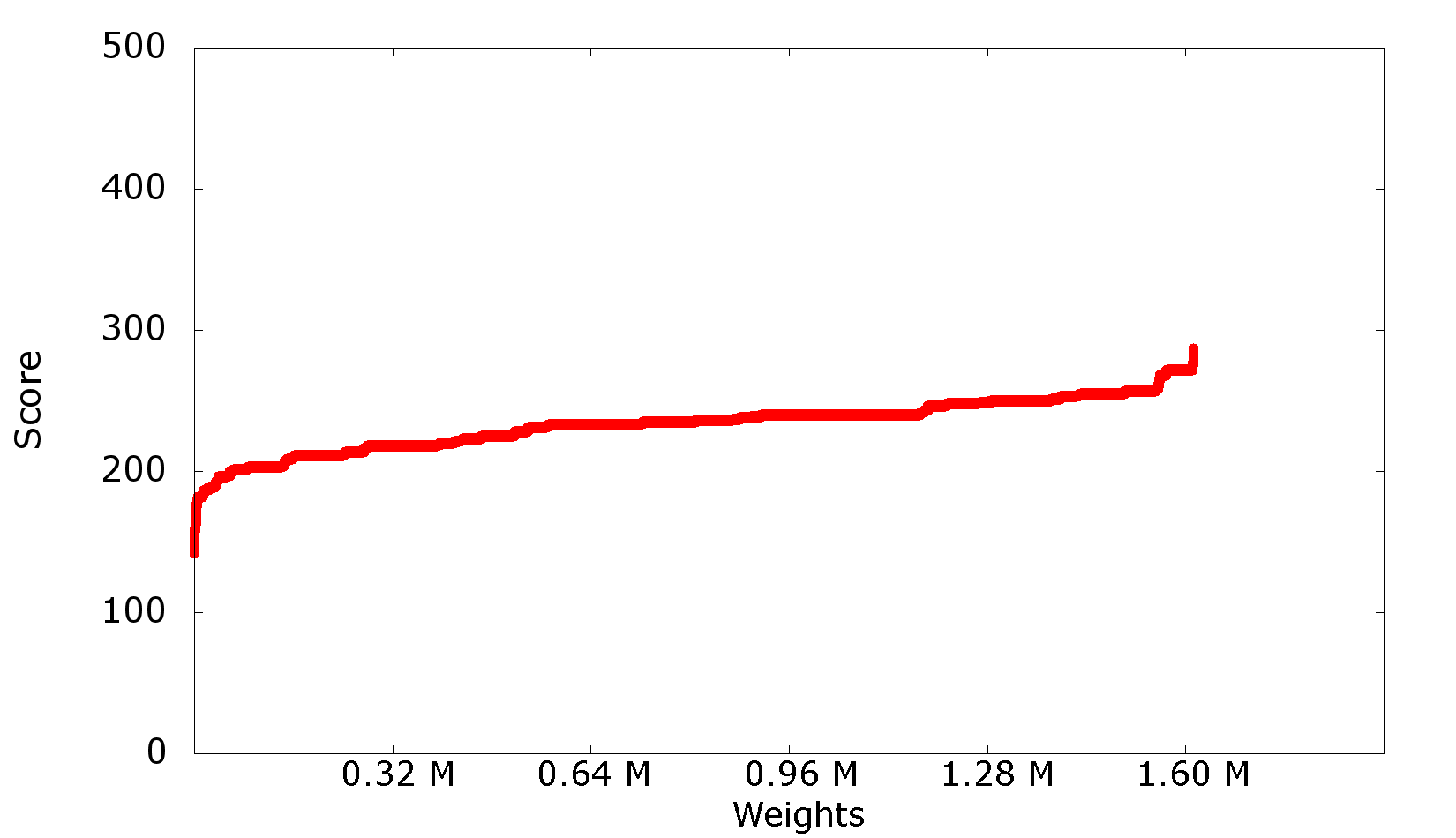} \label{figurec400-6}} \hfill
	\subfigure[Sequential PC]{\includegraphics[width=0.238\textwidth]{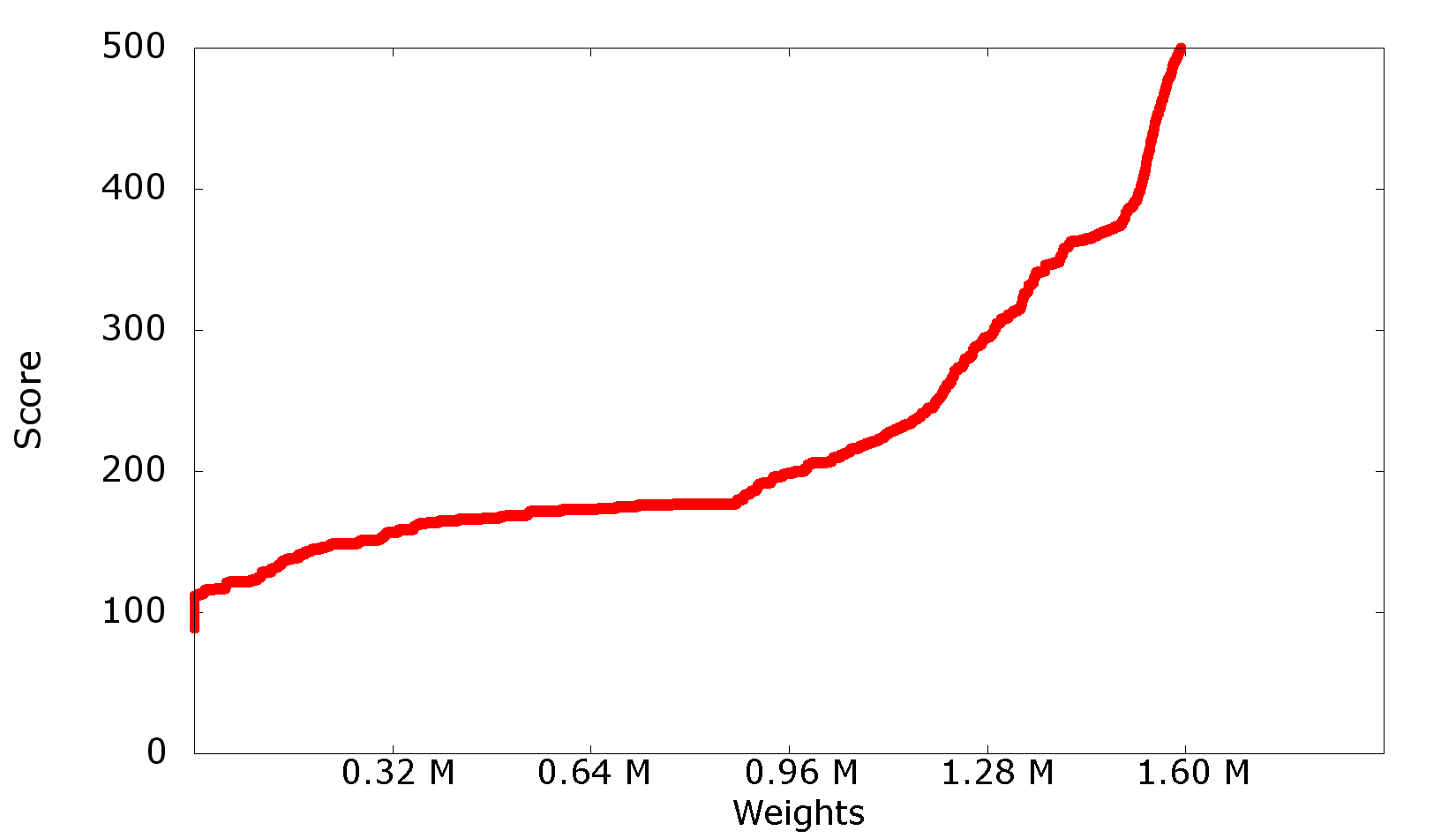} \label{figurec400-7}} \hfill
	\subfigure[Parallel PC]{\includegraphics[width=0.238\textwidth]{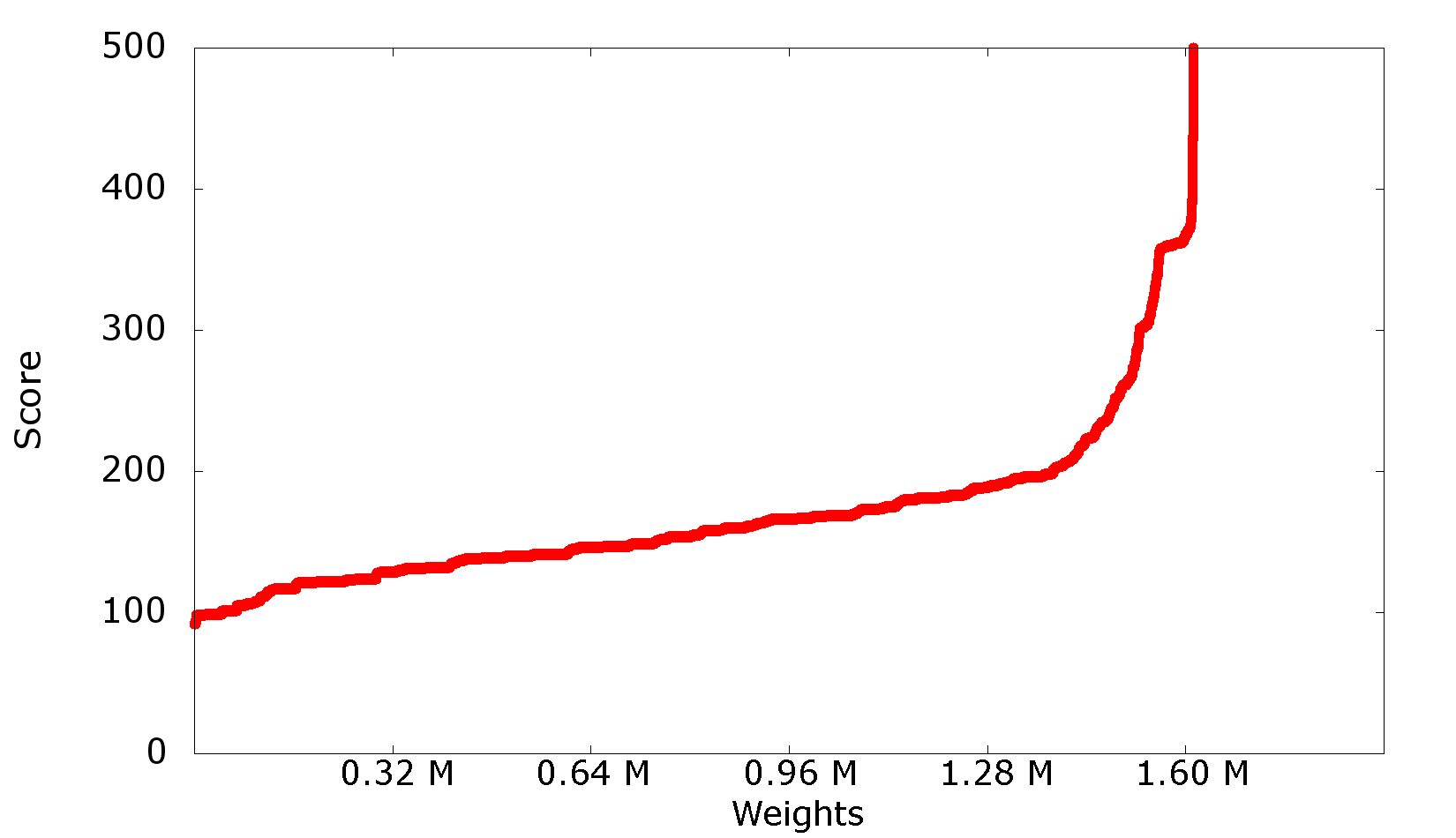} \label{figurec400-8}} \\
	\caption{Case Study 2 - Aggregate weights are shown in Figure \ref{figurec400-1} to Figure \ref{figurec400-4} and fine-grain weights are shown in Figure \ref{figurec400-5} to Figure \ref{figurec400-8}. X-axis is the weights ordered by scores shown in the Y-axis.}
	\label{figurec400}
\end{center}
\end{figure*}

\begin{figure*} 
\begin{center}
	\subfigure[Sequential P]{\includegraphics[width=0.238\textwidth]{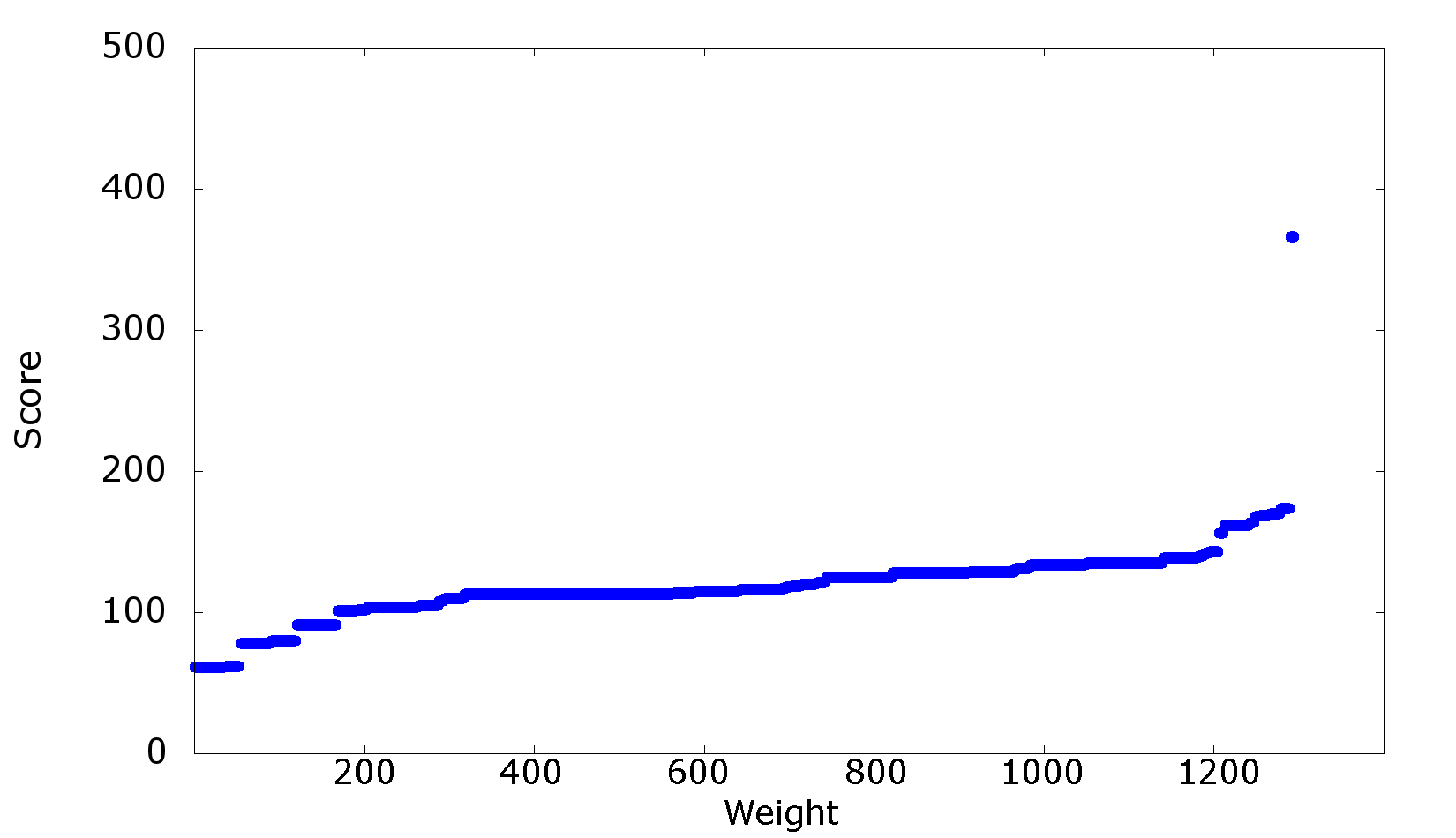} \label{figurec500-1}} \hfill
	\subfigure[Parallel P]{\includegraphics[width=0.238\textwidth]{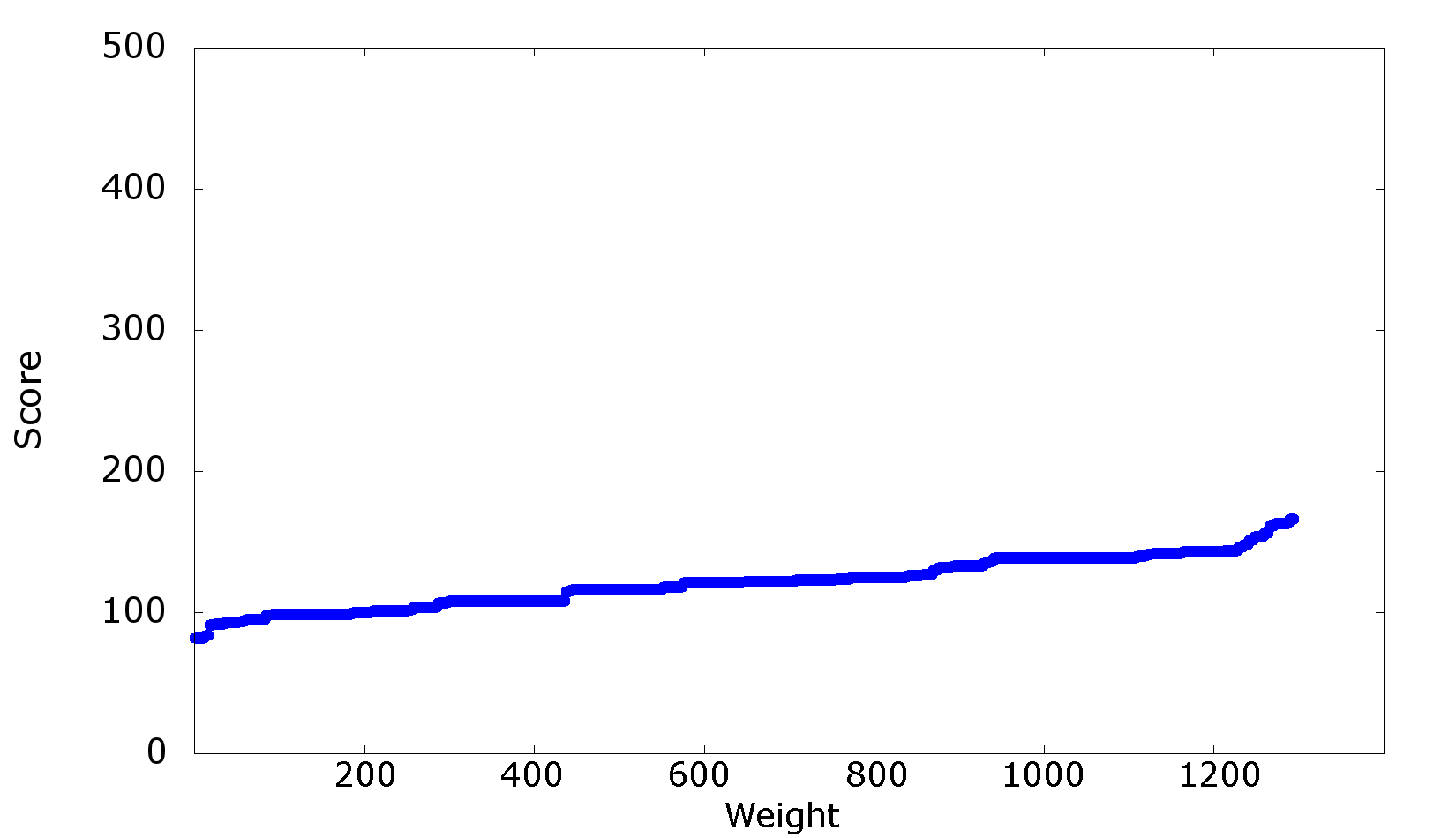} \label{figurec500-2}} \hfill
	\subfigure[Sequential PC]{\includegraphics[width=0.238\textwidth]{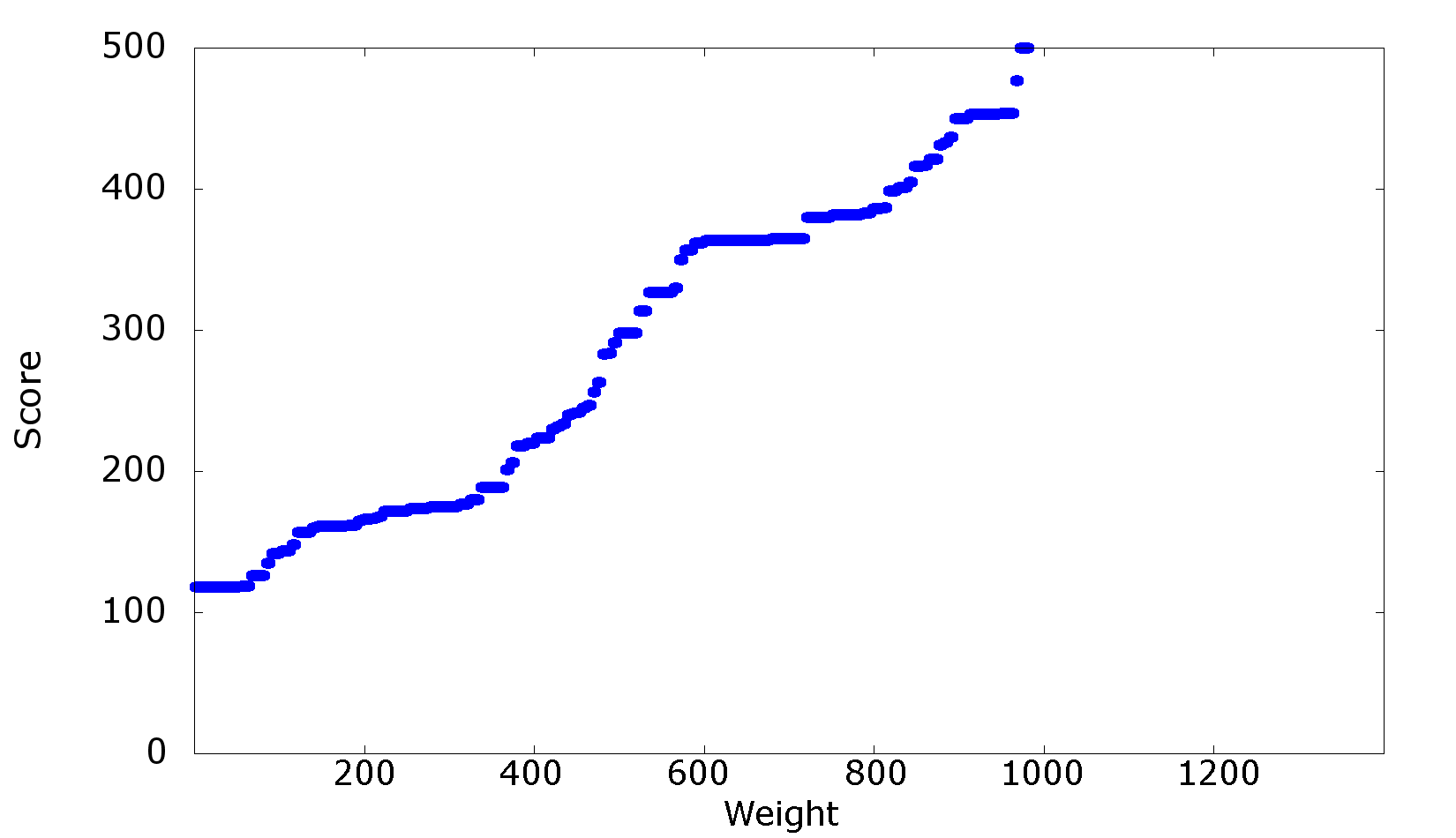} \label{figurec500-3}} \hfill
	\subfigure[Parallel PC]{\includegraphics[width=0.238\textwidth]{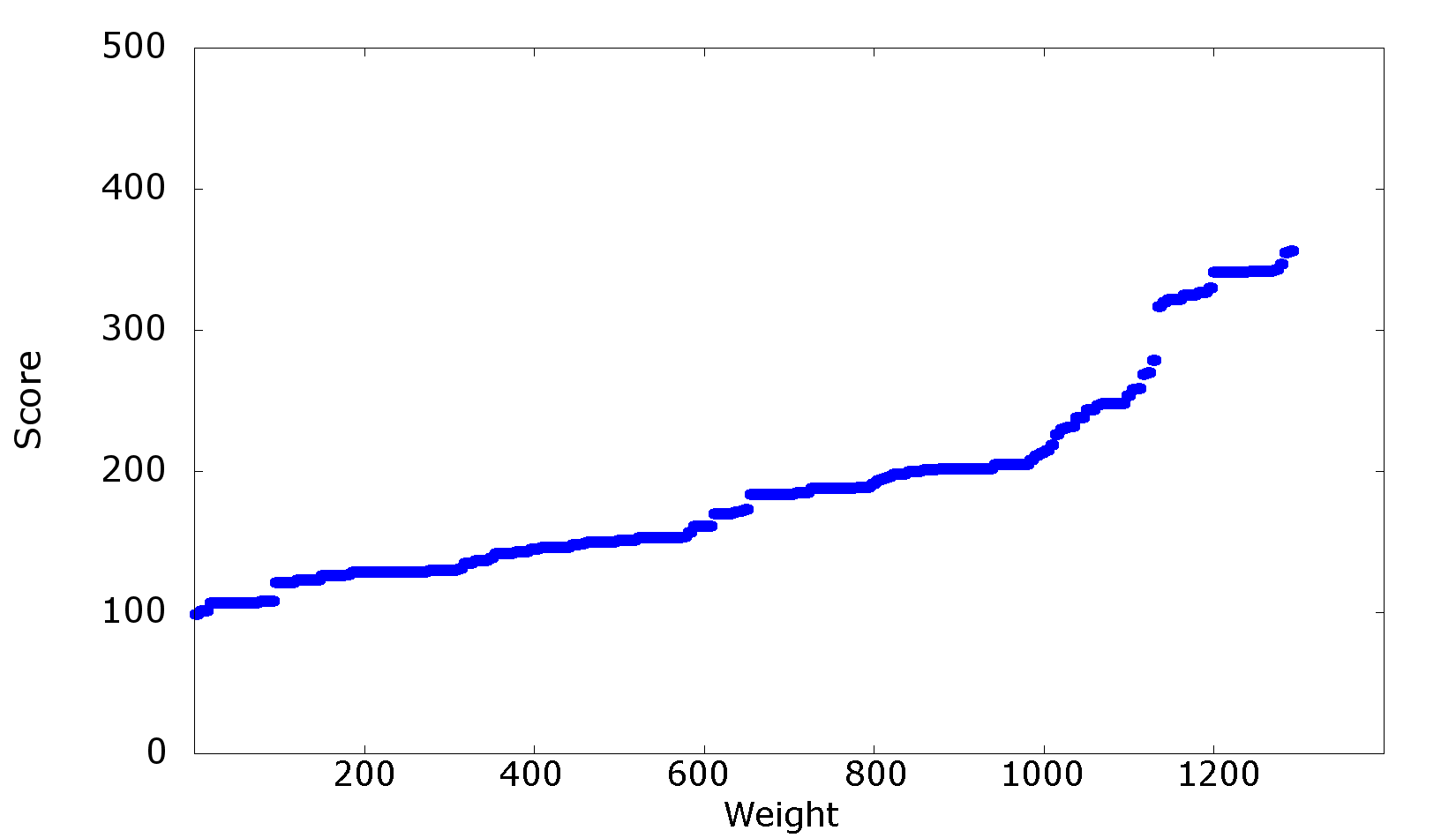} \label{figurec500-4}} \\
	\subfigure[Sequential P]{\includegraphics[width=0.238\textwidth]{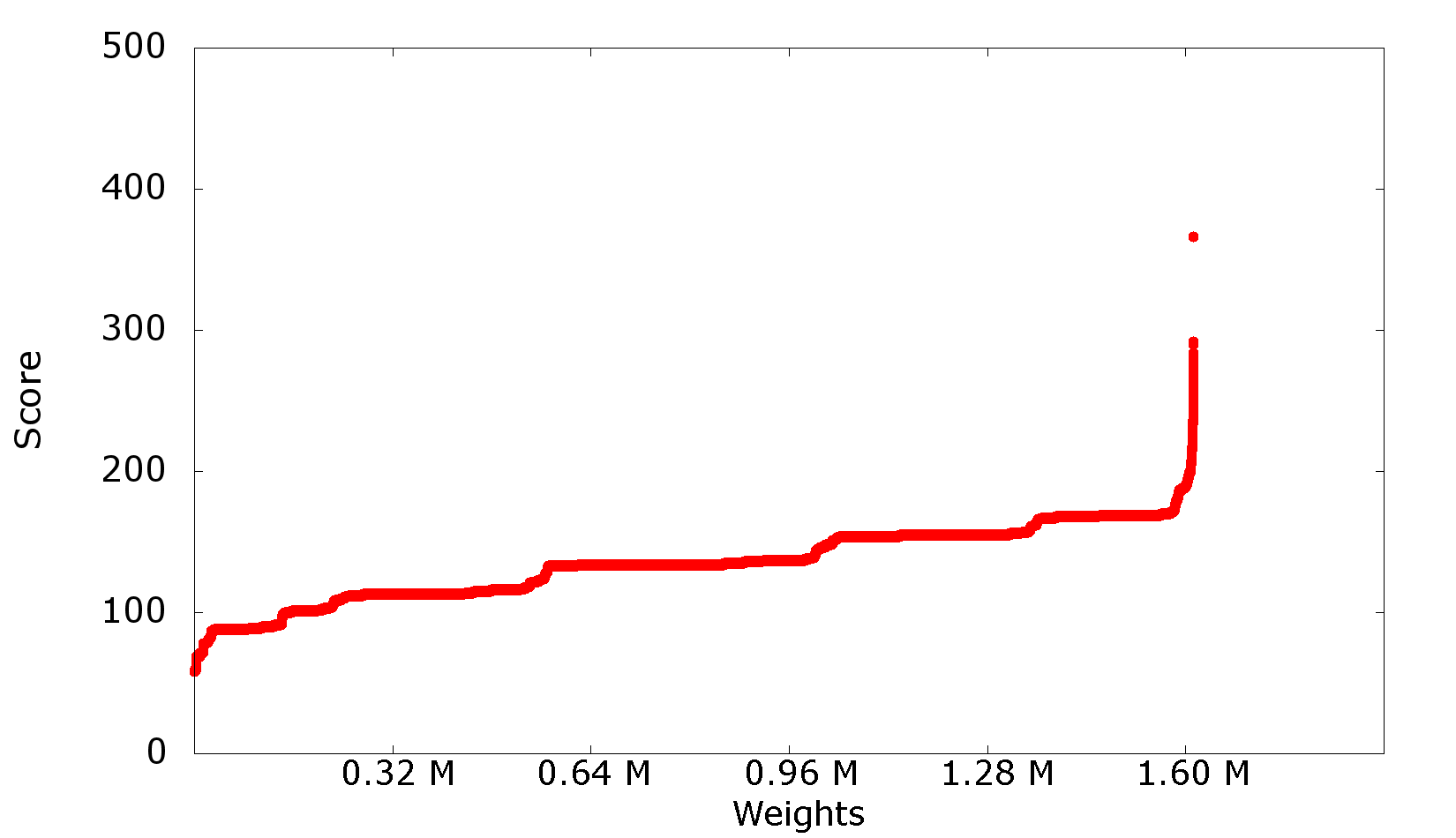} \label{figurec500-5}} \hfill
	\subfigure[Parallel P]{\includegraphics[width=0.238\textwidth]{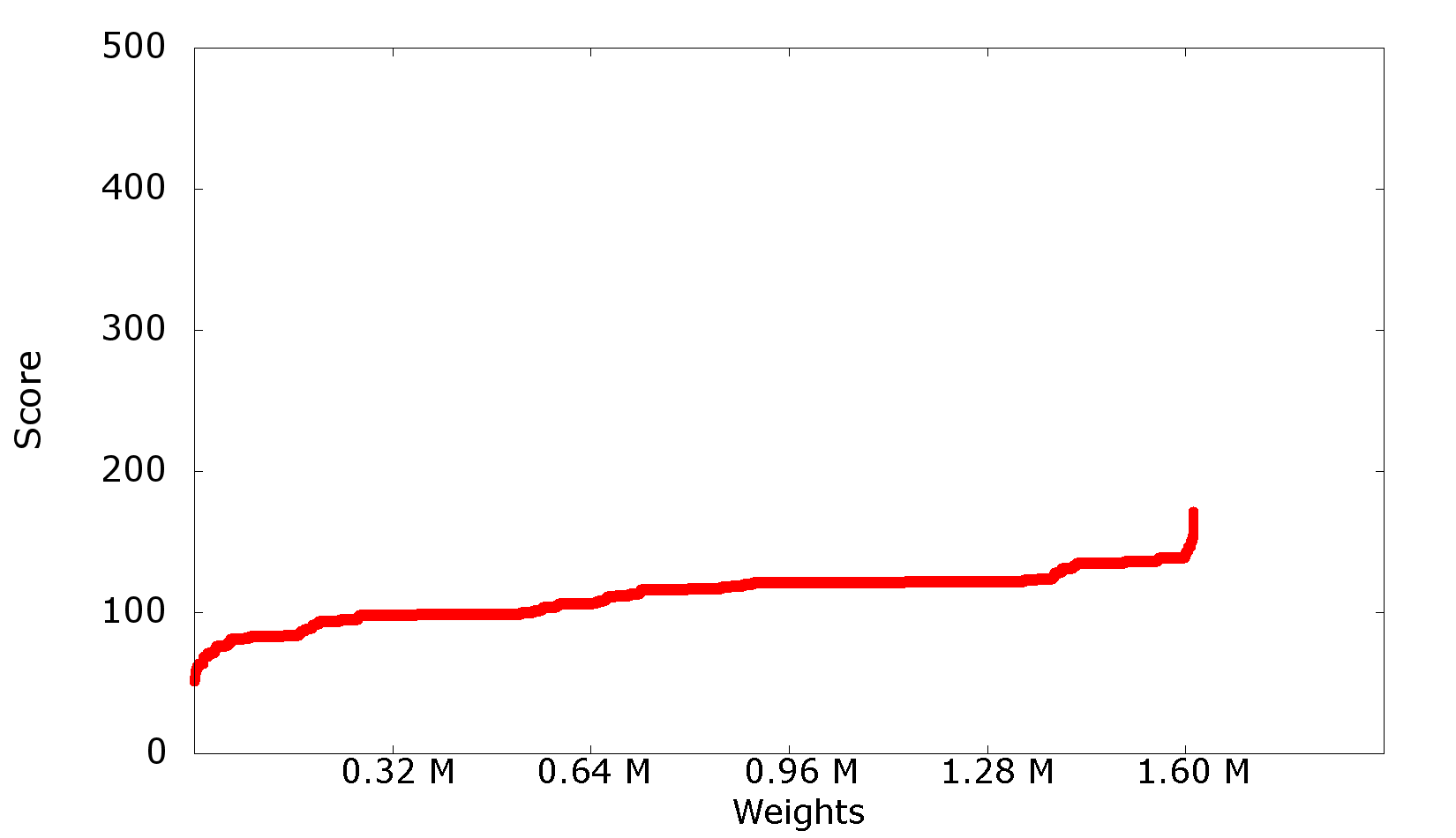} \label{figurec500-6}} \hfill
	\subfigure[Sequential PC]{\includegraphics[width=0.238\textwidth]{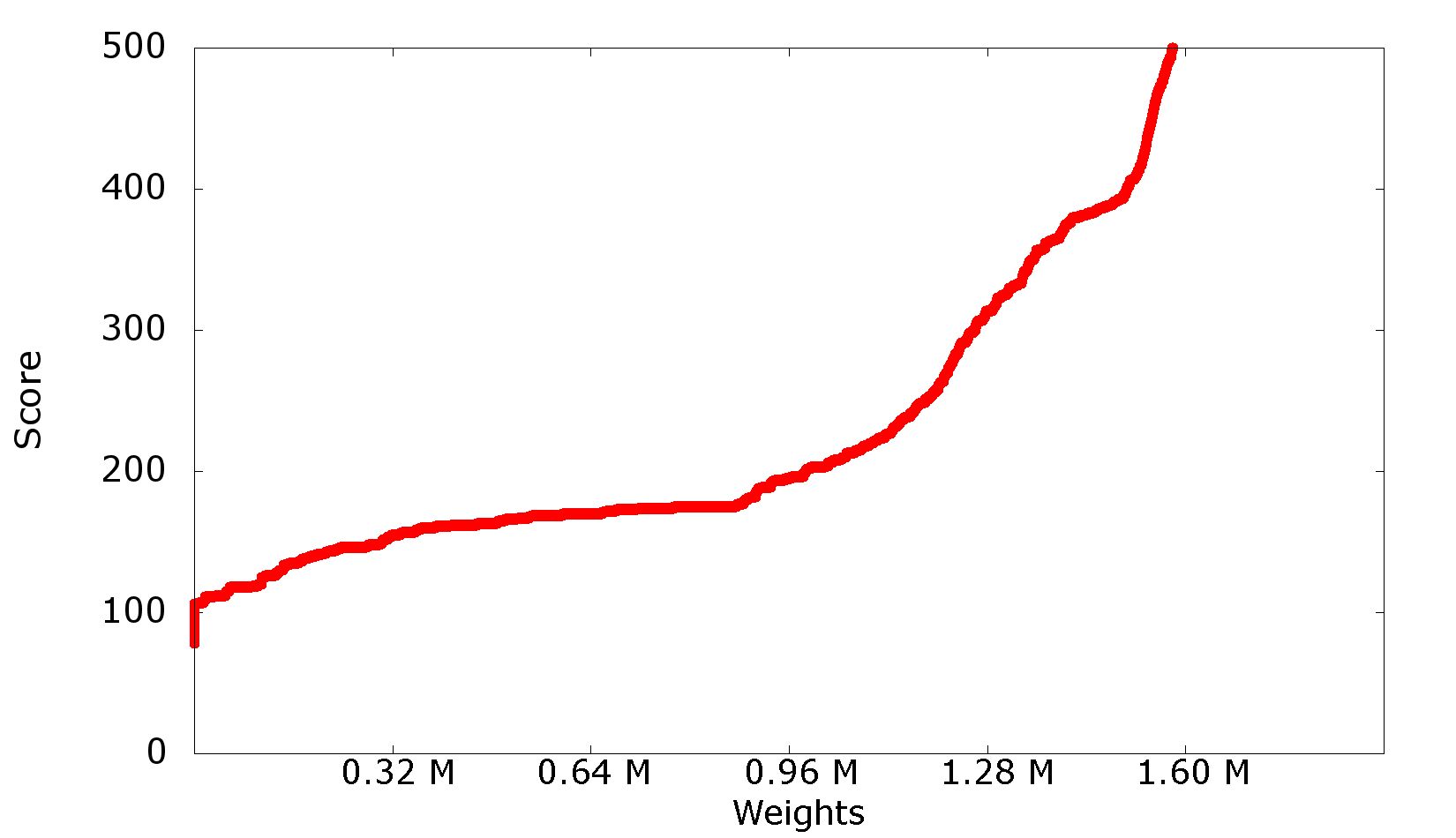} \label{figurec500-7}} \hfill
	\subfigure[Parallel PC]{\includegraphics[width=0.238\textwidth]{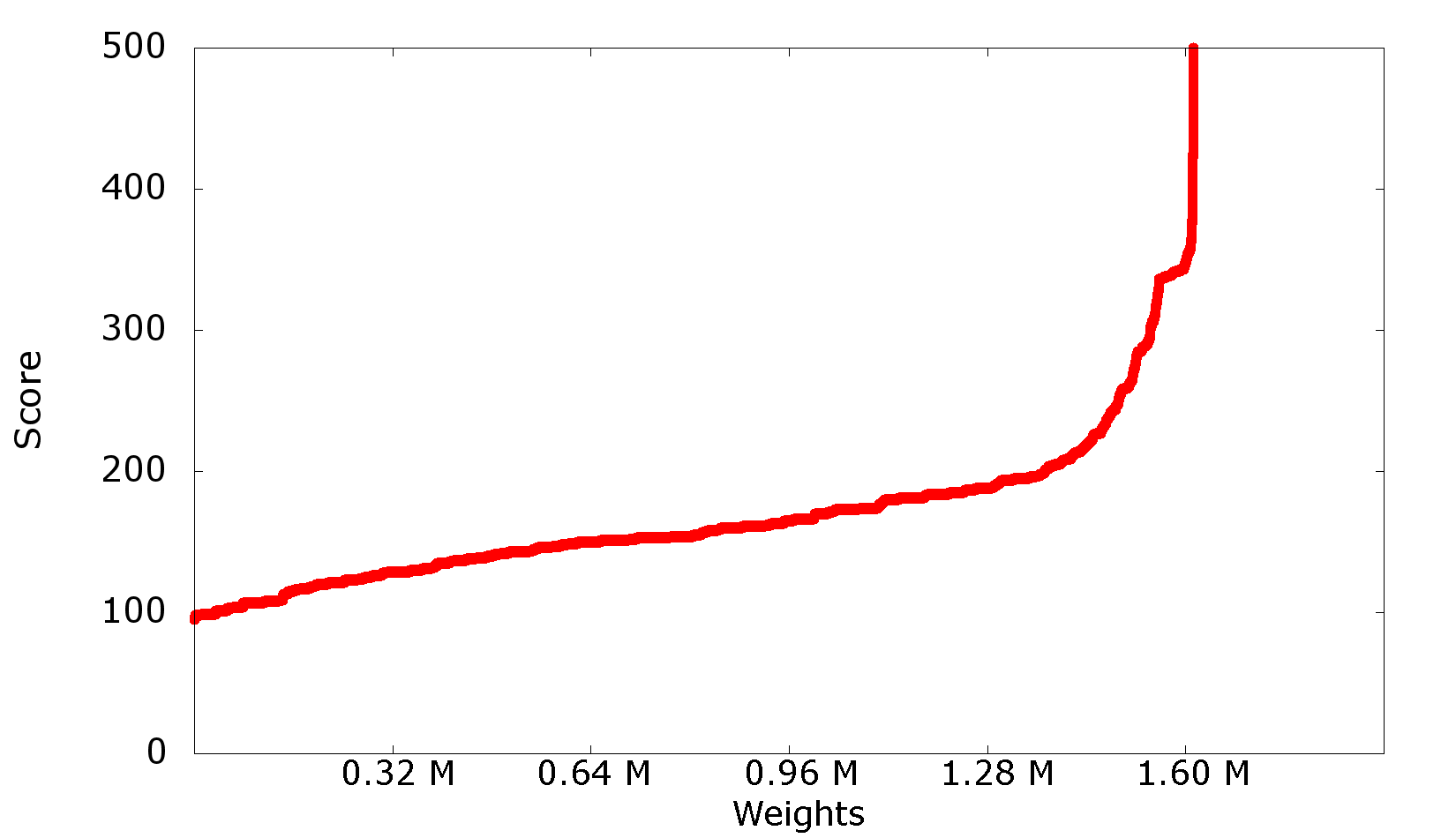} \label{figurec500-8}} \\
	\caption{Case Study 3 - Aggregate weights are shown in Figure \ref{figurec500-1} to Figure \ref{figurec500-4} and fine-grain weights are shown in Figure \ref{figurec500-5} to Figure \ref{figurec500-8}. X-axis is the weights ordered by scores shown in the Y-axis.}
	\label{figurec500}
\end{center}
\end{figure*}

Figure \ref{figurec300}, Figure \ref{figurec400} and Figure \ref{figurec500} show the space of different sets of weights and the scores of their rankings on the Y-axis. The X-axis are the different weights sorted by the scores. We make the following observations: (i) different set of aggregate weights produce rankings of distinct quality. This evidences that weights can discriminate between good and bad performance of VMs, (ii) the weights are more significant when costs are taken into account together with performance. The scores of the ranks are more diverse resulting in a discrimination of the weight space, (iii) using fine-grain weights results in increased number of options that are not available using aggregate weights, and (iv) fine-grain weights can produce better ranking options than aggregate weights.


\begin{table}
	\centering
	\caption{Percentage correlation of rankings obtained using fine-grain weights from benchmarking methodology and empirical analysis}
	\begin{tabular}{| p{1.6cm} | p{1.1cm} | p{1.1cm} | p{1.1cm} | p{1.1cm} |}
	\hline
	Case Study	& Sequential P rank	& Parallel P rank	& Sequential PC rank	& Parallel PC rank\\
	\hline
	1	&	93	&	87	&	93	&	91	\\
	2	&	85	&	67	&	96	&	97	\\
	3	&	81	&	95	&	95	&	95	\\
	\hline
	\end{tabular}
	\label{correlationtable2}
\end{table}

\section{Related Work}
\label{relatedwork}
Benchmarking is performed using a set of standard tests for evaluating the relative performance of a computing resource \cite{cloudbenchmark-5, cloudbenchmark-6}. For example, Linpack is used to evaluate the performance of supercomputers for ranking the Top500\footnote{\url{http://www.top500.org/}} list \cite{linpack-1}. Similar techniques can be employed to benchmark the cloud \cite{benchmark-2, cloudbenchmark-7, cloudbenchmark-8}. Cloud benchmarking considers the evaluation of the resources and the services \cite{cloudbenchmark-9, cloudbenchmark-9a}.

Unlike systems like the cluster or grid, the user has access to an abstract computing resource (limited access to the underlying hardware) on the cloud. Hence, it is important to understand the VM through resource benchmarking for obtaining maximum performance when an application is deployed. Performance of the memory accessible on the VM, processor capabilities such as computations performed, and file operations are taken into account. Such benchmarks are useful when a user wants to exploit multiple CPUs offered by cloud VMs.

Service benchmarking provides insight into the reliability and variability of cloud services \cite{cloudbenchmark-10}. For example, network performance between cloud resources is crucial for workflows or web services \cite{cloudbenchmark-1, cloudbenchmark-11, cloudbenchmark-12, cloudbenchmark-12a, cloudbenchmark-12b, cloudbenchmark-12c}. Sometimes, a hybrid of resource and service benchmarking is considered \cite{cloudbenchmark-13, cloudbenchmark-14}. In this paper, we assume that reasonable and seamless service is obtained on the public cloud and therefore only resource benchmarking is considered. 

Current resource benchmarking research on the cloud is limited in three ways. Firstly, research reported in relevant literature considers a small sample of low-cost VMs \cite{cloudbenchmark-3, cloudbenchmark-4}. In reality, a wide variety of VMs with varying performance and cost are available from a cloud provider which are not taken into account. The cloud infrastructure has rapidly matured over the last few years and consequentially, there has been a significant improvement in the performance of the VMs offered on the cloud. In this paper, we have considered a variety of resources with different performance capabilities offered by the same provider.

Secondly, the requirements of applications that need to be deployed on the cloud are seldom mapped on to benchmarks (for example, \cite{cloudbenchmark-10}). Benchmarking without meaningfully interpreting the results based on the requirements of an application cannot be useful to a user \cite{benchmark-2, cloudbenchmark-8}. In this paper, the proposed benchmarking methodology considers the requirements of the application and the user assigns weights of importance to cloud attributes that best describes the requirements. The result is a set of VMs on which maximum performance can be achieved.    

Thirdly, the benchmarking techniques need to incorporate methods to validate the benchmarks. This is an important issue to guarantee that the benchmarks obtained are acceptable. While a number of other issues related to benchmarking are addressed there is minimal focus in this direction (for example, \cite{cloudbenchmark-13, cloudbenchmark-4}). Empirical analysis is a straightforward way for validating benchmarks. While research reported in \cite{cloudbenchmark-17, cloudbenchmark-18, cloudbenchmark-19, cloudbenchmark-20, cloudbenchmark-20a, cloudbenchmark-21, cloudbenchmark-22} perform empirical analysis on the cloud, the results are not employed for validating the benchmark results. In this paper, we employ a validation technique in which the benchmarks are validated through an empirical analysis using case study applications. 


The research in this paper was motivated to address the above three challenges. For this, the cloud benchmarking methodology was developed, and in this research we have demonstrated the methodology on a set of resources with different performance capabilities, taken into account user requirements of an application, and validated the benchmarks using sample applications. The memory and process, local communication, computation and storage requirements of the application are known beforehand. These high-performance computing applications are embarrassingly parallel and executed on the multiple cores of a single VM. 

The state-of-the-art is advanced by developing an application aware benchmarking methodology that captures a wide variety of requirements of an application to generate performance and cost based rankings of cloud VMs. The methodology has been rigorously validated in multiple ways for three scientific HPC applications. In contrast to our previous paper \cite{cloudbenchmark-0}, the research presented in this paper improves on the benchmarking methodology by (i) generating both performance and cost based rankings, (ii) capturing the requirements of an application in a fine-grain manner using a set of eight weights, and (iii) validating the weight space using an enumeration technique, all of which were not considered previously.    

\section{Conclusions}
\label{conclusions}
In this paper, we proposed a six step benchmarking methodology for cloud VMs. We hypothesized that by considering the requirements of an application provided by a user as a set of weights (aggregate weights or fine-grain weights), along with benchmarking data collected from the cloud using the proposed methodology, VMs can be ranked in order of performance and cost effectiveness so that a user can deploy an application on a cloud VM, which will maximise performance. The hypothesis was validated on three real-world case study applications, namely a financial risk simulation, a molecular dynamics simulation and a mathematical solver, using comparative validation and enumeration-based validation techniques. The results show a good correlation between benchmarked and empirical rankings. 

The class of problems we have targeted in this paper are scientific HPC applications. They are usually developed and maintained by a large community and there is a lot of knowledge on their requirements. As an application evolves over time some requirements may vary, but these are generally known to the developers. The weights we determined for the applications were in consultation with experts and developers from each of these communities which was readily known to them. 

The benchmarking methodology we proposed can be used until a user is satisfied with the rankings generated. Once benchmarking is performed the entire weight space and rankings can be generated. The user can then provide different sets of weights (if not sure about certain weights) to obtain different rankings. Then the user can simply select the VMs which may appear in all the top ranks for different weights provided to ensure that performance is maximised.

Benchmarking the whole VM with large memory and storage is only possible at the expense of monetary costs and time. The benchmarking method proposed in this paper is limited in that way since it benchmarks a whole VM and thereby cannot be used in real-time. However, the benchmark data obtained from the method can be used with real-time methods. We are currently investigating container as a mechanism for benchmarking a small portion of a VM to reduce the time and costs required for cloud benchmarking to bypass the above limitation. Preliminary experiments indicate that container-based benchmarking can give similar results to benchmarking the whole VM. Using this method saves both time and money spent for benchmarking. 

In the future, we intend to incorporate benchmarking useful for applications that span across multiple nodes and leverage multi-core systems with complex memory hierarchies, such as GPUs.

\ifCLASSOPTIONcompsoc
  \section*{Acknowledgments}
\else
  \section*{Acknowledgment}
\fi

This research was pursued under the EPSRC grant, EP/K015745/1, a Royal Society Industry Fellowship and an AWS Education Research grant.  

\ifCLASSOPTIONcaptionsoff
  \newpage
\fi

\begin{IEEEbiography}[{\includegraphics[width=1in,clip,keepaspectratio]{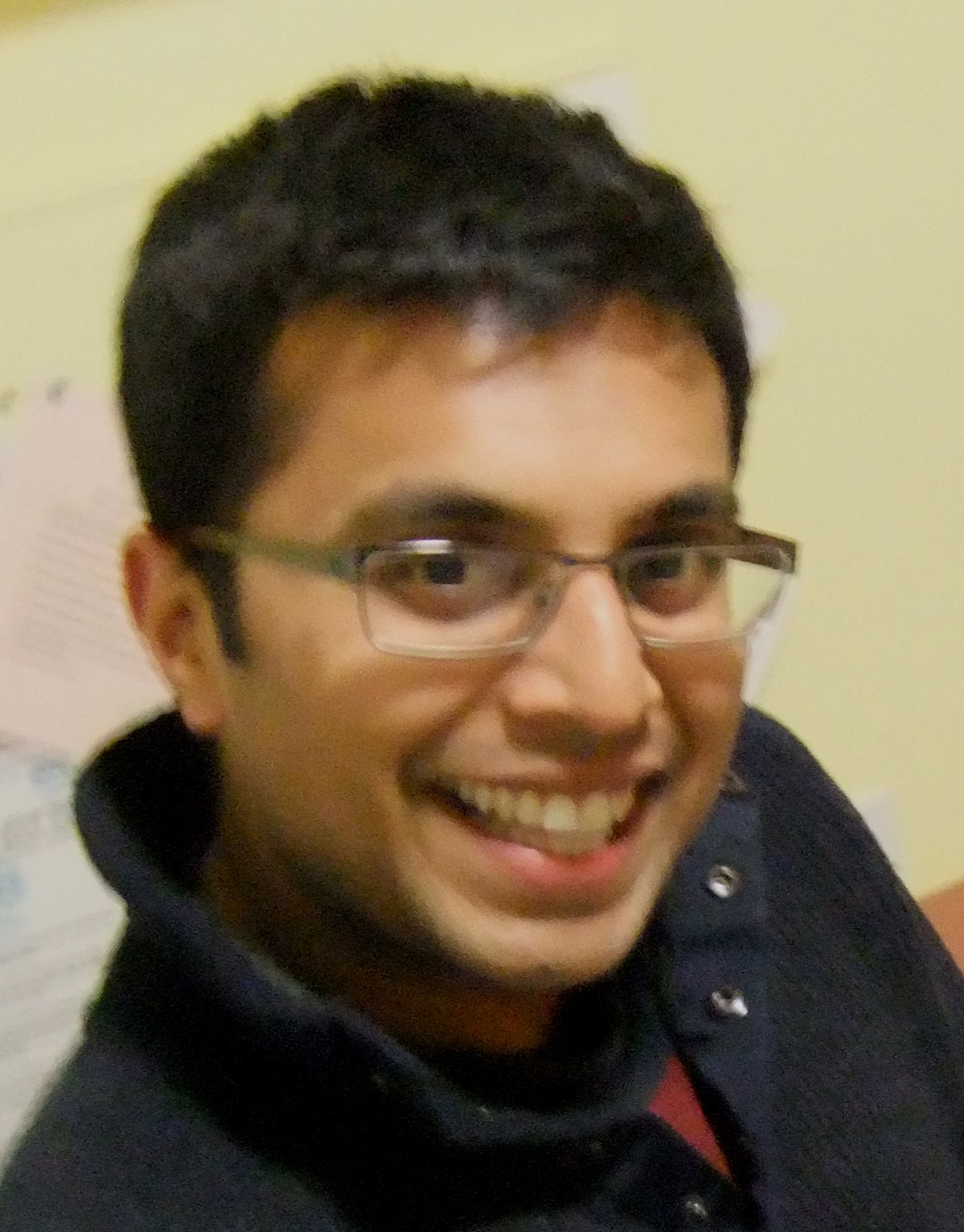}}]{Blesson Varghese}
is a Research Fellow in Computer Science at the Queen's University Belfast. Previously, he worked at University of St Andrews and Dalhousie University. He obtained a PhD in Computer Science (2011) and MSc in Network Centred Computing (2008), both from the University of Reading, on international scholarships. His research interests are in developing and analysing parallel and distributed systems. More information is available from www.blessonv.com. 
\end{IEEEbiography}

\begin{IEEEbiography}[{\includegraphics[width=1in,clip,keepaspectratio]{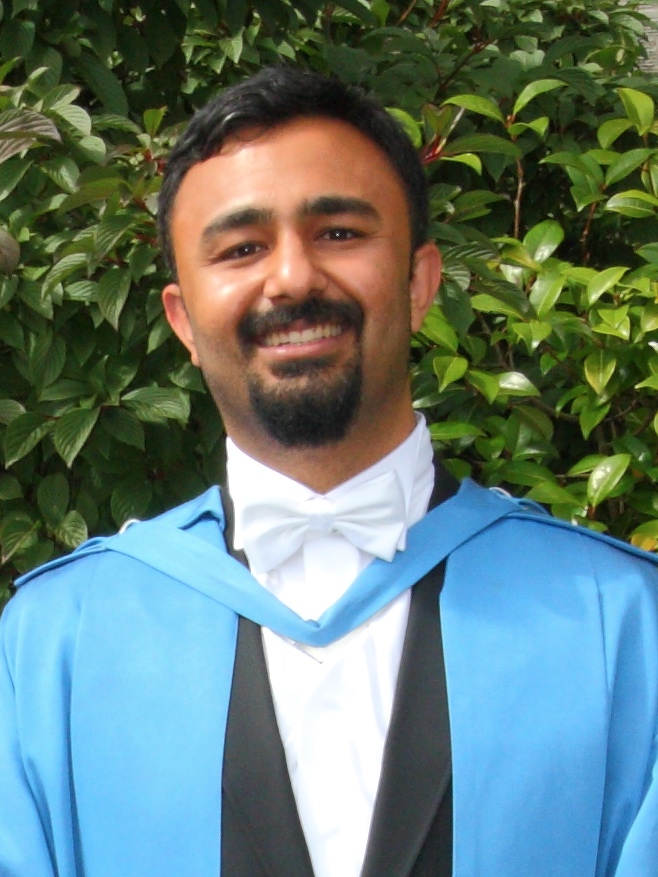}}]{Ozgur Akgun}
is a Research Associate at the School of Computer Science, University of St Andrews. He received BSc in Computer Science from the Izmir University of Economics (2009) and PhD in Computer Science from the University of St Andrews (2014). His research focus is on high level modelling languages for constraint programming and their use for automated modelling. He is interested in applications of constraint programming in cloud computing and astronomy.
\end{IEEEbiography}

\begin{IEEEbiography}[{\includegraphics[width=1in,clip,keepaspectratio]{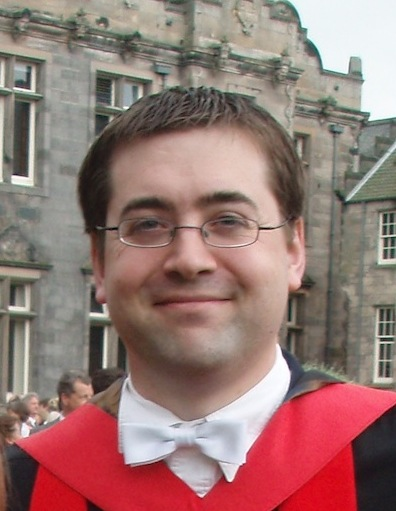}}]{Ian Miguel}
is a professor of Computer Science at the University of St Andrews. He has researched Constraint Programming throughout his career. His PhD received the BCS/CHPC Distinguished Dissertation award, and between 2004 and 2009 he held a Royal Academy of Engineering/EPSRC Research Fellowship. Recently, he has focused on automated constraint modelling and constructing efficient constraint solvers. Ian is Principal Investigator of UK EPSRC EP/K015745/1, having previously been PI or CI of four other EPSRC grants and an EPSRC CASE for New Academics award, sponsored by Microsoft Research, totalling nearly GBP 3M.
\end{IEEEbiography}

\begin{IEEEbiography}[{\includegraphics[width=1in,clip,keepaspectratio]{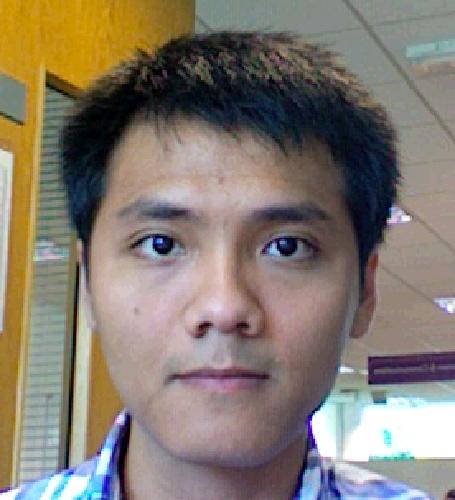}}]{Long Thai}
is a PhD student at School of Computer Science, St Andrews University. He received his BEng in Information Technology from Ho Chi Minh City International University, and his MSc in Internet and Distributed Systems from Durham University. His research focuses on resource provisioning and job scheduling on the cloud to satisfy performance and monetary cost requirements. 
\end{IEEEbiography}

\begin{IEEEbiography}[{\includegraphics[width=1in,clip,keepaspectratio]{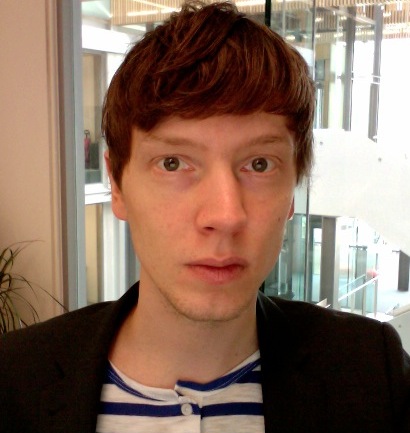}}]{Adam Barker}
is a Senior Lecturer in Computer Science at the University of St Andrews, a Royal Society Industry Fellow, and an Honorary Fellow at the University of Edinburgh. He worked as a Research Fellow at the University of Oxford, University of Melbourne and University of Edinburgh. He holds a PhD in Informatics from the University of Edinburgh (2007). Adam's primary research interests concentrate on the effective engineering of large-scale distributed systems, covering areas such as cloud computing, big data infrastructure, data-intensive computing and service-oriented architecture. 
\end{IEEEbiography}

\end{document}